\documentclass[aps,prd,nobibnotes,twocolumn,superscriptaddress,bibliography]{revtex4-2}
\usepackage{amsfonts}
\usepackage{mathrsfs}
\usepackage{amsmath}
\usepackage{color}
\usepackage{graphicx}
\usepackage{bm}
\usepackage{amssymb}
\usepackage{xspace}
\usepackage{epstopdf}
\usepackage{dcolumn}
\usepackage{float}
\usepackage{comment}
\usepackage{braket}
\usepackage{soul}
\usepackage{multirow}
\usepackage{longtable}
\usepackage{xcolor}

\usepackage{scrextend}

\usepackage[colorlinks=true, letterpaper=true, pdfstartview=FitV, linkcolor=blue, citecolor=blue, urlcolor=blue]{hyperref}

\makeatother
\begin{document}
	\title{Spin layer groups and their co representations}

\author{Zeying Zhang}
\thanks{These authors contributed equally to this work}
\email{zzy@mail.buct.edu.cn}
\affiliation{College of Mathematics and Physics, Beijing University of Chemical
	Technology, Beijing 100029, China}

\author{Gui-Bin Liu$^*$}
\email{ gbliu@bit.edu.cn}
\affiliation{Centre for Quantum Physics, Key Laboratory of Advanced Optoelectronic Quantum Architecture and Measurement (MOE),
	Beijing Institute of Technology, Beijing 100081, China}

\author{Mu Tian}
\affiliation{Centre for Quantum Physics, Key Laboratory of Advanced Optoelectronic Quantum Architecture and Measurement (MOE),
	Beijing Institute of Technology, Beijing 100081, China}

\author{Run-Wu Zhang}
\affiliation{Centre for Quantum Physics, Key Laboratory of Advanced Optoelectronic Quantum Architecture and Measurement (MOE),
	Beijing Institute of Technology, Beijing 100081, China}

\author{Zhi-Ming Yu}
\affiliation{Centre for Quantum Physics, Key Laboratory of Advanced Optoelectronic Quantum Architecture and Measurement (MOE),
	Beijing Institute of Technology, Beijing 100081, China}

\author{Yugui Yao}
\email{ygyao@bit.edu.cn}
\affiliation{Centre for Quantum Physics, Key Laboratory of Advanced Optoelectronic Quantum Architecture and Measurement (MOE),
	Beijing Institute of Technology, Beijing 100081, China}

\begin{abstract}
Spin layer groups are the crystallographic symmetry groups  with a periodic plane, and their symmetry operations are inherited from three-dimensional (3D) spin space groups. However, the direct application of 3D symmetry groups to two-dimensional systems is often inadequate due to anisotropic axes and dimensional reduction. In this work, we systematically classify inequivalent spin layer groups and analytically derive their irreducible corepresentations. This classification establishes a foundational framework for investigating symmetry-protected properties and novel quantum states in low-dimensional magnetic materials.
\end{abstract}
\maketitle
	
\section{Introduction}
Symmetry plays a central role in condensed matter physics, with group theory providing its fundamental mathematical language. When time-reversal symmetry is incorporated into crystalline symmetries, the resulting groups are known as magnetic groups \cite{dresselhaus_group_2008,bradley_mathematical_2009,litvin_magnetic_2016}. These groups extend the ordinary point groups or space groups and enable the description of a wide range of magnetic phenomena. 
In magnetic groups, spin rotations are locked to lattice rotations due to spin–orbit coupling (SOC).
Breakng this lock makes spin and lattice can rotate independently, leading to the formation of more complex symmetry groups known as spin groups \cite{w_f_brinkman_theory_1966, litvin_spin_1974, litvin_spin_1977}. Therefore, spin groups are particularly suitable for describing magnetic systems with negligible  or weak SOC, such as those composed of light elements or low-angular-momentum electrons. The spin-group extension of space group is called spin space group (SSG).
Recently, three-dimensional (3D) SSGs have been systematically investigated and classified into collinear, coplanar, and noncoplanar types, each potentially hosting different physical properties \cite{chen_enumeration_2024,jiang_enumeration_2024,xiao_spin_2024}. Based on spin groups, novel phenomena have been predicted systematically. Specifically, 218 novel quasiparticles unique to spin groups have been identified \cite{liu_spin-group_2022},  498 unconventional magnon modes have been reported \cite{chen_unconventional_2025}. {\color{black}More importantly, altermagnetism has been identified as a novel magnetic class that exhibits substantial momentum-dependent spin splitting in the absence of SOC, which is naturally explained by the collinear spin groups \cite{smejkal_beyond_2022, jungwirth_altermagnetism_2025}.} 

In recent years, intrinsic two-dimensional (2D) magnetic materials have been synthesized, such as CrI$_3$, Cr$_2$Ge$_2$Te$_6$, and Fe$_3$GeTe$_2$ \cite{huang_layer-dependent_2017,gong_discovery_2017,deng_gate-tunable_2018}. These systems provide a versatile platform for exploring emergent quantum phenomena. {\color{black}Notably, the altermagnetic phase is not restricted to 3D bulk crystals but can be naturally inherited by 2D systems. Within the spin group framework, these low-dimensional magnets maintain the characteristic spin splitting.} For example, tunable altermagnetism has been engineered in twisted van der Waals magnets \cite{liu_twisted_2024}, while spin-valley locking has been reported in layered altermagnets \cite{ablimit_weak_2018}. Moreover, spin groups reveal exotic quasiparticles in low-dimensional magnets, including chiral Dirac fermions in CoNb$_3$S$_6$ \cite{ghimire_large_2018}, eightfold degenerate points in 2D magnetic systems \cite{guo_eightfold_2021} and engineering band structure in 2D square tessellations \cite{che_engineering_2025}. These developments highlight the growing importance of spin group symmetries in describing unconventional magnetic phases, particularly in low-dimensional systems.

From a theoretical perspective, the reduced dimensionality of 2D systems imposes restriction on magnetic materials. According to the Mermin-Wagner theorem, long-range magnetic order in two dimensions requires strong magnetic anisotropy to suppress thermal fluctuations \cite{mermin_absence_1966}. In 2D materials, such anisotropy can arise from both SOC and crystal field effects \cite{gong_two-dimensional_2019,wang_magnetic_2022}, reflecting the intricate interplay between spin and lattice degrees of freedom. More importantly, the reduction from three to two dimensions breaks the equivalence of spatial directions and introduces anisotropic axes, leading to a restructuring of symmetry operations.

As a consequence, while 3D SSGs can in principle be applied to 2D systems by incorporating a sufficiently thick vacuum layer. However,  they do not provide a natural or minimal description of the underlying symmetries. In a 3D framework, symmetry operations irrelevant to the 2D geometry are formally retained, leading to redundancy and a lack of adaptation to reduced dimensionality. To address these issues, one must introduce spin layer groups (SLGs), which extend ordinary layer groups within the spin-group framework. Similar to the case in conventional crystallography, where the 80 layer groups can be mapped to the 230 space groups \cite{kopsky_international_2002}.  The symmetry operations of SLGs can be inherited from 3D SSGs, the correspondence is not one-to-one due to dimensional reduction and anisotropy, such that distinct SLGs may map to the same SSG. Consequently, a systematic classification of SLGs is essential for establishing a complete and non-redundant symmetry description of 2D magnetic materials. While the concept of SLGs has been applied in the study of 2D altermagnetism~\cite{ zeng_description_2024}, a comprehensive classification and the associated irreducible corepresentations (coreps) remain lacking, limiting symmetry-based analysis and prediction of physical phenomena in low-dimensional magnetic systems.

In this work, we systematically classify inequivalent SLGs and analytically derive their irreducible coreps, thereby providing a complete symmetry framework for two-dimensional magnetic systems.

\section{Construction of SLG\lowercase{s}}

Generally, there are two fundamental approaches to constructing SLGs.
The first proceeds from the 80 layer groups by associating them with irreducible real representations of the orthogonal group $O(N)$ $(N=1,2,3)$ and enumerating their normal subgroups or supergroups \cite{jiang_enumeration_2024,chen_enumeration_2024,xiao_spin_2024}.
The second approach is based on dimensional reduction, recognizing that the symmetry operations of SLGs descend naturally from three-dimensional spin space groups. Specifically, SLGs are obtained by restricting 3D SSGs to systems with periodicity in two dimensions \cite{zhang_encyclopedia_2023}.
In this work, we adopt the latter approach and classify all inequivalent SLGs within this framework.


\subsection{Dimensional reduction}

Generally, consider an SLG $\mathcal{L}$.  A corresponding SSG $\mathcal{S}$  can be constructed from $\mathcal{L}$ as follows. Let $\mathcal{N}$ denote the lattice translation group normal to the periodic plane of $\mathcal{L}$. The SSG is then obtained as the semi-direct product 
\begin{equation}\label{eq:L}
	\mathcal{S}=\mathcal{N}\rtimes \mathcal{L}.
\end{equation}
Since $\mathcal{N}$ is a normal subgroup of $\mathcal{S}$, we have 
\begin{equation*}
	\mathcal{L}\cong \mathcal{S}/ \mathcal{N}.
\end{equation*}
The corresponding isomorphic map $\phi:\mathcal{L}\rightarrow\mathcal{S}/ \mathcal{N}$ is defined by
\[
\phi: \{S||R|t\}\mapsto \{S||R|\tau\}\mathcal{N},
\]
where $\{S||R|t\}\in \mathcal{L}$, $\{S||R|\tau\}\in \mathcal{S}$, and the translation part $t$  is given by the projection of  $\tau$  onto the periodic plane.
The normality of $\mathcal{N}$ imposes a constraint on the spatial rotation part $R$ of $\mathcal{S}$. For $\{E||E|t_3\}\in \mathcal{N}$ and $\{S||R|\tau\}\in \mathcal{S}$, one has
\begin{equation}
	\label{eq:norsub}
	\{S||R|\tau\}\{E||E|t_3\}\{S||R|\tau\}^{-1}=\{E||E|Rt_3\}\in \mathcal{N}.
\end{equation}
This requires $Rt_3=\pm t_3$, i.e., each rotation $R$ preserves both the 2D subspace $V_2$ of the periodic plane and the one-dimensional (1D) subspace $V_1$  orthogonal to it. 
Accordingly, the matrix representation of $R$ takes a block-diagonal form $O(2) \oplus O(1)$, where $O(2)$ acts on $V_2$  (leaving $t_3$ invariant) and $O(1) = \{\pm1\}$ acts on $V_1$ (along the direction of $t_3$).

The above facts provide a practical route to construct SLGs from SSGs. Starting from a given SSG $\mathcal{S}$, one identifies 1D translation subgroups $\mathcal{N}$ that are normal in $\mathcal{S}$, and then constructs the corresponding SLG through the quotient $\mathcal{S}/\mathcal{N}$.
In practice, for each SSG, one can generate symmetry-equivalent realizations by cyclic permutations of the coordinate axes $a,b,c \rightarrow b,c,a \rightarrow c,a,b$, yielding corresponding axes $a',b',c'$. 
For each such realization, one examines whether the 1D translation subgroup along the $c'$-direction forms a normal subgroup of $\mathcal{S}$ according to Eq.~\eqref{eq:norsub}. If this condition is satisfied, an SLG can be obtained from $\mathcal{S}$.
The resulting SLGs may still be equivalent under origin shifts or axis transformations. The final step is therefore to identify inequivalent SLGs.

\subsection{Equivalence of SLGs}

In general, two SSGs $\mathcal{S}_1$ and $\mathcal{S}_2$ are considered equivalent if and only if there exist an origin shift $\tau$, a lattice transformation $Q$, and a rotation $P$ such that \cite{jiang_enumeration_2024,chen_enumeration_2024}
\begin{equation*}
	\{S^\prime||R^\prime|t^\prime\}
	=\{PSP^{-1}||Q RQ^{-1}|Qt-QRQ^{-1}\tau+\tau\},
\end{equation*}
where $\{S||R|t\}\in\mathcal{S}_1$, $\{S'||R'|t'\}\in\mathcal{S}_2$, $\tau\in\mathbb{R}^3$, $Q\in SL(3,\mathbb{Z})$, and $P\in O(3)$.
For SLGs, however, the inequivalence between lattice vectors in the periodic plane and those normal to it imposes additional constraints on $\tau$ and $Q$. In particular, the lattice transformation $Q$ must preserve the decomposition of space into the two subspaces $V_2$ and $V_1$.  
Accordingly, $Q$ takes a block-diagonal form $Q = Q_2 \oplus Q_1$, 
where $Q_2 \in SL(2,\mathbb{Z})$ acts within the periodic plane and $Q_1 \in SL(1,\mathbb{Z})=\{\pm1\}$ acts along the normal direction. Correspondingly, the origin shift satisfies $\tau \in \mathbb{R}^2$.
For two SLGs $\mathcal{L}_1$ and $\mathcal{L}_2$, one first determines whether there exist $Q$ and $\tau$ such that
\begin{equation}
	\label{eq:equ}
	\begin{cases}
		R^\prime=Q RQ^{-1},\\
		t^\prime=Qt-QRQ^{-1}\tau+\tau,
	\end{cases}
\end{equation}
for all $\{S||R|t\}\in \mathcal{L}_1$ and $\{S^\prime||R^\prime|t^\prime\}\in \mathcal{L}_2$. 
If no such $Q$ and $\tau$ exist, the two SLGs are inequivalent. Otherwise, one must further check whether the spin-rotation parts can be matched by a real orthogonal matrix $P$. If all these conditions are satisfied, $\mathcal{L}_1$ and $\mathcal{L}_2$ are regarded as equivalent.
Eq.~\eqref{eq:equ} constitutes a system of linear congruence equations modulo 1. These equations can be solved efficiently using the Smith decomposition \cite{cohen_course_1993, eick_computing_1997}, as described in Appendix~\ref{app:lce}.

\begin{table}[t]
	\caption{Summary of SLGs with $I_k \leq 12$. Here $N$ denotes the number of inequivalent SLGs determined by Eq.~\eqref{eq:equ}, and $N_{3D}$ the number of inequivalent SSGs that admit a decomposition $\mathcal{S}=\mathcal{L}\ltimes \mathcal{N}$. The quantities $N_1$, $N_2$, and $N_3$ denote the numbers of SSGs that correspond to one, two, and three inequivalent SLGs, respectively. Accordingly, $N = N_1 + 2N_2 + 3N_3$ and $N_{3D} = N_1 + N_2 + N_3$.}
	\label{tab:slg}
	\begin{ruledtabular}
		\begin{tabular}{llllll}
			SLGs &$N$& $N_{3D}$ & $N_1$& $N_2$& $N_3$ \\
			\hline
			Collinear & 448 &398&349&48&1\\
			Coplanar&6083 &5457&4839&610&8\\ 
			Non-coplanar &33556&30599&27670&2901&28\\
			Total &40087&36454&32858&3559&37
		\end{tabular}
	\end{ruledtabular}
\end{table}

In Ref.~\cite{jiang_enumeration_2024}, SSGs are constructed from a space group $G$ and its normal subgroup $H$. Two indices, $I_k$ and $I_t$, are used to label SSGs. Let $T_G$ and $T_H$ denote the translation subgroups of $G$ and $H$, respectively, and $P_G$ and $P_H$ their corresponding point groups. Then $I_k$ is defined as $|T_G/T_H|$, reflecting the supercell multiplicity,  and $I_t$ is the ratio of point-group orders, $I_t = |P_G|/|P_H|$.
Here, we extend the notation in Ref.~\cite{jiang_enumeration_2024} to label SLGs:
\[
N_{SG}.I_k.I_t.N_{\mathrm{rep}}(.L(P)).l
\]
where $N_{SG}$ denotes the space group number, and $N_{\mathrm{rep}}$ labels the corresponding 3D representation (3D) for given $I_k$ and $I_t$. The symbol $.L(P)$ distinguishes collinear and coplanar SSGs. The index $l$ (with maximal value 3) labels SLGs that correspond to the same 3D SSG $N_{SG}.I_k.I_t.N_{\mathrm{rep}}(.L(P))$ but are inequivalent in two dimensions.
Starting from 1421 collinear, 24,788 coplanar-noncollinear, and 157,289 noncoplanar SSGs, we obtain 448 collinear, 6083 coplanar-noncollinear, and 33,556 noncoplanar SLGs with $I_k \leq 12$. The numbers of SLGs in each category are summarized in Table~\ref{tab:slg}.
In most cases, a given SSG corresponds to a unique SLG. However, multiple inequivalent SLGs may correspond to the same SSG due to dimensional reduction. For example, the collinear SSG $25.1.2.2.L$ gives rise to three inequivalent SLGs, namely $25.1.2.2.L.1$, $25.1.2.2.L.2$, and $25.1.2.2.L.3$. This multiplicity highlights the necessity of distinguishing SLGs when analyzing 2D magnetic systems.
Detailed correspondences between SLGs and SSGs are provided in Sections S1--S3 of the Supplemental Material \cite{noauthor_notitle_nodate}.

\section{Structure of SLG\lowercase{s}}
\label{sec:SLGstruct}

Every SLG can be expressed as a direct product of a nontrivial SLG $\mathcal{L}^\prime$ and a spin-only group $\mathcal{S}_\mathrm{so}$, with a suitable choice of $\mathcal{L}^\prime$ \cite{litvin_spin_1974}. Here, a nontrivial SLG refers to one without pure spin rotations other than the identity, while a spin-only group consists solely of pure spin rotations. That is,
\begin{equation}
\label{eq:str}
\mathcal{L}=\mathcal{L}^\prime\times \mathcal{S}_\mathrm{so}.
\end{equation}
In fact, this structure is not specific to SLGs, but applies more generally to spin groups. Here we just describe it in the context of SLGs.
Without loss of generality, we take the magnetization direction to be along the $z$-axis for collinear configurations, and within the $xy$ (periodic) plane for coplanar ones (throughout this paper, “coplanar” refers to coplanar-noncollinear configurations). Then $\mathcal{S}_\mathrm{so}$ for different magnetic configurations can be written as
\begin{equation}
\label{eq:Sso}
\mathcal{S}_\mathrm{so}=\begin{cases}
	\mathcal{C}_{\infty}+\{C_{2 x}\mathcal{T}||E|\boldsymbol{0}\}\mathcal{C}_\infty & \text{collinear},\\
	\{ \{E||E|\boldsymbol{0}\},\{C_{2 z}\mathcal{T}||E|\boldsymbol{0}\} \} & \text{coplanar},\\
	\{ \{E||E|\boldsymbol{0}\} \} & \text{noncoplanar},
\end{cases}
\end{equation}
where $\mathcal{C}_\infty=\{\{C_{\theta z}||E|\boldsymbol{0}\}\mid \theta\in [0,2\pi)\}\cong SO(2)$ and $\mathcal{T}$ is the time-reversal operation.  The nontrivial part $\mathcal{L}'$ may be unitary or not,
\[
\mathcal{L}^\prime=\begin{cases}
	\mathcal{U}^\prime & \text{unitary},\\
	\mathcal{U}^\prime + \mathcal{A}^\prime \mathcal{U}^\prime & \text{non-unitary},
\end{cases}
\]
where $\mathcal{U}'$ is the maximal unitary subgroup of $\mathcal{L}'$ and $\mathcal{A}'$ is an anti-unitary element in $\mathcal{L}'$.

For collinear SLGs, the structure of $\mathcal{L}^\prime$ exhibits additional simplifications and can be related directly to ferromagnetic (FM), antiferromagnetic (AFM), or altermagnetic (AM) configurations.
In collinear situation, the spin-rotation part of $\mathcal{L}^\prime$, denoted as $\mathcal{S}_0^\prime$, is $\{E\}$ for unitary $\mathcal{L}'$ and is $\{E, \mathcal{T}\}$ for non-unitary $\mathcal{L}'$. 
When $\mathcal{S}_0^\prime=\{E\}$, no operation in $\mathcal{L}'$ reverses the spins (magnetic moments), corresponding to FM case. 
In contrast, when $\mathcal{S}_0^\prime=\{E, \mathcal{T}\}$,  $\mathcal{A}'$ can take the form of 
\begin{equation}
  \label{eq:Ap}
  \mathcal{A}'=\mathcal{T}U'_0=\{\mathcal{T}||R'_0|t'_0\},
\end{equation}
where $U'_0=\{E||R'_0|t'_0\}$ is an unitary operation that exchanges sublattices with opposite spins and $\mathcal{T}$ reverses the spins simultaneously, 
corresponding to AFM or AM cases.

\section{Corepresentations of SLG\lowercase{s}}

In principle, the coreps of SLGs can be obtained either from little-group theory or from those of the corresponding SSGs.
Specifically, one approach is to compute the small coreps of the little groups of SLGs, from which all coreps can be constructed by induction \cite{liu_msgcorep_2023}. 
Alternatively, the coreps of an SLG $\mathcal{L}$ can be derived from the coreps $D$ of an SSG $\mathcal{S}$ satisfying $\mathcal{S}=\mathcal{L}\ltimes \mathcal{N}$. In this case, the relevant coreps are those satisfing $\ker D \supseteq \mathcal{N}$ in which 
$\ker D=\{g\in \mathcal{S}\mid D(g)=I\}$ \cite{james_representations_2001,zhang_encyclopedia_2023}.
While the coreps of SSGs have been computed in previous works \cite{jiang_enumeration_2024,chen_enumeration_2024,xiao_spin_2024,song_constructions_2025}, there remain two main limitations. First, existing databases are constructed using Hamiltonian-based methods, and the results are largely numerical \cite{song_constructions_2025}, which hinders further analytical treatment. Second, in Refs.~\cite{song_constructions_2025,chen_enumeration_2024,xiao_spin_2024}, no publicly accessible database is available.
To overcome these limitations, we employ little-group theory to derive all coreps of SLGs analytically \cite{bradley_mathematical_2009}.

\subsection{Little group of SLG}

The little group at $k$ point of $\mathcal{L}$, denoted as $\mathcal{L}_k$, is defined by:
\begin{equation}
\mathcal{L}_k=\{\{S||R||t\} \in \mathcal{L} \mid Rk=(-1)^{\nu(S)}k+K_m\},
\end{equation}
where $K_m$  is a reciprocal lattice vector, $\nu(S)=0$ if $S$ is a unitary operator and $\nu(S)=1$ if $S$ is a anit-unitary operator. The method to find all  high-symmetry $k$  points is elaborated  in Appendix~\ref{app:findk}. 
The maximal unitary subgroup of  $\mathcal{L}_k$ is denoted as
\[
\mathcal{U}_k=\{\{S||R|t\} \in \mathcal{L }_k \mid \nu(S)=0\}.
\]
Furthermore, $\mathcal{L}_k$ is treated differently depending on whether it's unitary
\begin{equation}
\label{eq:LkUk}
\mathcal{L}_k=\begin{cases}
\mathcal{U}_k & \text{unitary} \\ 
\mathcal{U}_k+\mathcal{A}\mathcal{U}_k & \text{non-unitary}
\end{cases}\
\end{equation}
where $\mathcal{A}$  is an anti-unitary operator (not necessarily identical to $\mathcal{A}'$). The key to constructing the small coreps of $\mathcal{L}_k$ is to obtain the small reps of $\mathcal{U}_k$.
The structure of $\mathcal{U}_k$ depends on the magnetic type of an SLG.


\subsubsection{$\mathcal{U}_k$ of collinear SLG}

When $\mathcal{L}'=\mathcal{U}'$, i.e. the collinear FM case, considering the collinear $\mathcal{S}_\mathrm{so}$ in Eq. \eqref{eq:Sso}, the little group $\mathcal{L}_k$ is 
\[
\begin{split}
\mathcal{L}_k=&
[\mathcal{U}^\prime]_k\mathcal{C}_\infty+\{C_{2 x}\mathcal{T}||E|\boldsymbol{0}\}[\mathcal{U}^\prime]_{-k}\mathcal{C}_\infty\\
\end{split},
\]
where $[\mathcal{U}']_k$ and $[\mathcal{U}']_{-k}$ are defined as follows
\[
\begin{split}
   [\mathcal{U}']_k&=\{\{S||R|t\}\in\mathcal{U}' \mid Rk=k+K_m\}, \\
   [\mathcal{U}']_{-k}&=\{\{S||R|t\}\in\mathcal{U}' \mid Rk=-k+K_m\}.
\end{split}
\]
The first term of the above $\mathcal{L}_k$ is its unitary subgroup $\mathcal{U}_k$.
If  $[\mathcal{U}']_{-k} = \varnothing$ then $\mathcal{L}_k=\mathcal{U}_k$.
Otherwise, $\mathcal{L}_k=\mathcal{U}_k+\mathcal{AU}_k$ and any element in $\{C_{2 x}\mathcal{T}||E|\boldsymbol{0}\}[\mathcal{U}^\prime]_{-k}\mathcal{C}_\infty$  can be chosen as $\mathcal{A}$, e.g. $\mathcal{A}=\{C_{2x}\mathcal{T}||R|t\}$ in which $\{E||R|t\}\in[\mathcal{U}']_{-k}$.
Since every element of $\mathcal{U}'$ commutes with every element of $\mathcal{C}_\infty$. we have a direct product structure of $\mathcal{U}_k$
\[
\mathcal{U}_k=[\mathcal{U}^\prime]_k\mathcal{C}_\infty=[\mathcal{U}^\prime]_k \times \mathcal{C}_\infty.
\]

When $\mathcal{L}'=\mathcal{U}'+\mathcal{A}'\mathcal{U}'$, i.e. the collinear AFM or AM cases, the little group $\mathcal{L}_k$ is more complex
\begin{equation}
	\label{eq:AFMLk}
\begin{split}
\mathcal{L}_k=&[\mathcal{U}^\prime]_k\mathcal{C}_\infty + \{C_{2 x}||E|\boldsymbol{0}\}[{U}^\prime_0\mathcal{U}^\prime]_{k}\mathcal{C}_\infty +\\
&  \{C_{2 x}\mathcal{T}||E|\boldsymbol{0}\}[\mathcal{U}^\prime]_{-k}\mathcal{C}_\infty + \mathcal{T} [U^\prime_0\mathcal{U}^\prime]_{-k}\mathcal{C}_\infty,\\
\end{split} 
\end{equation}
where $ \mathcal{T}{U}^\prime_0=\mathcal{A}^\prime$ [c.f. Eq. \eqref{eq:Ap}]. The first two terms constitute $\mathcal{U}_k$. 
Note that the first term is always nonempty because $[\mathcal{U}']_k$ contains identity at least, but $[U'_0\mathcal{U}']_k=\varnothing$, $[\mathcal{U}']_{-k}=\varnothing$, and $[U'_0\mathcal{U}']_{-k}=\varnothing$ maybe occur for all the other three terms.
Since $C_{2x}C_{\theta z}C_{2x}^{-1}=C_{-\theta z} \neq C_{\theta z}$ and $\mathcal{C}_\infty$ is a normal subgroup of $\mathcal{L}_k$, the structure of  $\mathcal{U}_k$  is a semi-direct product rather than a direct product:
\[
\begin{split}
&\mathcal{U}_k=([\mathcal{U}^\prime]_k+\{C_{2 x}||E|\boldsymbol{0}\}[{U}^\prime_0\mathcal{U}^\prime]_{k}) \ltimes \mathcal{C}_\infty.
\end{split}
\]

Combining the two cases of $\mathcal{L}'$, $\mathcal{U}_k$ can be written as the following general form
\begin{equation}
\label{eq:colUk}
   \mathcal{U}_k=\mathcal{U}'_k \ltimes \mathcal{C}_\infty,
\end{equation}
where $\mathcal{U}'_k$ and $\mathcal{C}_\infty$ are the non-trivial spin group and spin-only group of $\mathcal{U}_k$, respectively.
Accordingly, the non-trivial part $\mathcal{U}'_k$ is
\begin{equation}
 \label{eq:Ukp}
 \mathcal{U}'_k=\begin{cases}
   [\mathcal{U}']_k & \text{FM} \\
   [\mathcal{U}^\prime]_k+\{C_{2 x}||E|\boldsymbol{0}\}[{U}^\prime_0\mathcal{U}^\prime]_{k} & \text{A(F)M},
 \end{cases}
\end{equation}
where A(F)M means AFM or AM.
In the FM case, the semidirect product in Eq. \eqref{eq:colUk} reduces to a direct product.

\subsubsection{$\mathcal{U}_k$ of coplanar SLG}

For coplanar SLGs, the situation is considerably simpler since $\mathcal{S}_{\mathrm{so}}$ contains only two elements.
When  $\mathcal{L}^\prime=\mathcal{U}^\prime$, we have
\[
\begin{split}
	\mathcal{L}_k=&[\mathcal{U}^\prime]_k+\{C_{2 z}\mathcal{T}||E|\boldsymbol{0}\}[\mathcal{U}^\prime]_{-k}.
\end{split}
\]
Conversely, if $\mathcal{L}^\prime=\mathcal{U}^\prime + \mathcal{A}^\prime \mathcal{U}^\prime $, the little group is
\[
\begin{split}
\mathcal{L}_k=&[\mathcal{U}^\prime]_k+\{C_{2 z}||E|\boldsymbol{0}\}[U^\prime_1\mathcal{U}^\prime]_{k}+\\
&\{C_{2 z}\mathcal{T}||E|\boldsymbol{0}\}[\mathcal{U}^\prime]_{-k}+ \mathcal{T} [U^\prime_1\mathcal{U}^\prime]_{-k},
\end{split}
\]
where $ \mathcal{T}{U}^\prime_1=\mathcal{A}^\prime$ and ${U}^\prime_1$ is a unitary operation whose spin-rotation part may not be $E$. 
Therefore, for coplanar SLGs, we have
\[
\mathcal{U}_k=  
\begin{cases}
[\mathcal{U}^\prime]_k & \mathcal{L}^\prime=\mathcal{U}^\prime \\
[\mathcal{U}^\prime]_k+\{C_{2 z}||E|\boldsymbol{0}\}[U^\prime_1\mathcal{U}^\prime]_{k}& \mathcal{L}^\prime=\mathcal{U}^\prime + \mathcal{A}^\prime \mathcal{U}^\prime. \\
\end{cases}
\]

\subsubsection{$\mathcal{U}_k$ of noncoplanar SLG}

This is the simplest case because $\mathcal{S}_\mathrm{so}$ has only identity and hence $\mathcal{L}=\mathcal{L}'$. 
The little group is
\[
  \mathcal{L}_k = 
\begin{cases}
[\mathcal{U}^\prime]_k & \mathcal{L}^\prime=\mathcal{U}^\prime \\
[\mathcal{U}^\prime]_k+\mathcal{T}[U^\prime_1\mathcal{U}^\prime]_{-k}& \mathcal{L}^\prime=\mathcal{U}^\prime + \mathcal{A}^\prime \mathcal{U}^\prime, \\
\end{cases}
\]
and we simply have
\[
  \mathcal{U}_k = [\mathcal{U}^\prime]_k.
\]



\subsection{Small coreps of $\mathcal{L}_k$}
\label{subsec:corep}

The two cases of $\mathcal{L}_k$ in Eq. \eqref{eq:LkUk} require separate treatments in the construction of its small coreps.
Denote the $p$-th small rep of $\mathcal{U}_k$ as $D_k^p$, and denote the small corep of $\mathcal{L}_k$ derived from $D_k^p$ as $\Gamma_k^p$.
When $\mathcal{L}_k=\mathcal{U}_k$, we simply have $\Gamma_k^p=D_k^p$. 
When $\mathcal{L}_k=\mathcal{U}_k+\mathcal{AU}_k$,  $\mathcal{U}_k$ is an index-2 unitary subgroup of $\mathcal{L}_k$, and $\Gamma_k^p$ can be derived from  $D_k^p$ according to the corep type/case of $D_k^p$ (or $\Gamma_k^p$) via standard procedures \cite{bradley_mathematical_2009,liu_msgcorep_2023}.
The corep type $\tau_k^p$ of $D_k^p$ is determined by 
\begin{align}
  \tau_k^p&=\frac{1}{|\mathcal{U}_k|}\sum_{g\in \mathcal{A}\mathcal{U}_k}\chi_k^p(g^2) \nonumber\\
	 &=\frac{1}{|\mathcal{U}_k/\!\!/ T|}\sum_{g\in \mathcal{A}\mathcal{U}_k/\!\!/T}\chi_k^p(g^2)
	=\begin{cases}
	+ 1& \text{case(a)}\\
	- 1& \text{case(b)}\\
	0 &\text{case(c)}\\
\end{cases}
\end{align}
where $\mathcal{A}\mathcal{U}_k/\!\!/T$ means the set of coset representatives for the cosets of the translation group $T$.
In cases (a) and (b), one can find a matrix $N$ that makes $D_k^p(g)$  equivalent to $D_k^p(\mathcal{A}^{-1}g\mathcal{A})^*$. In case (c) $D_k^p(g)$  is not equivalent to $D_k^p(\mathcal{A}^{-1}g\mathcal{A})^*$.  The degeneracy exhibits fundamentally different behavior in each case:
in case (a), no additional degeneracy exists and in  both cases (b) (two copies of the same small reps) and (c) (two different small reps) have additional degeneracies.

\subsection{Small reps of $\mathcal{U}_k$}
 \label{subsec:rep}
The small reps of $\mathcal{U}_k$  can be obtained from the  projective reps of the little co-group $ \bar{\mathcal{U}}_k$ of $\mathcal{U}_k$
\[
\bar{\mathcal{U}}_k=\{\{S||R\} \mid \{S||R|t\} \in \mathcal{U}_k\}.
\]
Two group elements $g_i=\{S_i||R_i\}$ and $g_j=\{S_j||R_j\}$ in $\bar{\mathcal{U}}_k$ satisfy the following multiplication rule
\[
g_ig_j=\{S_iS_j||R_iR_j\}.
\]
For given $k$ point, the  projective rep matrices of $\bar{\mathcal{U}}_k$ satisfy  
\begin{equation}
	\bar D(g_i)\bar D(g_j)=\mu(g_i,g_j)\bar D(g_ig_j),
\end{equation}
where the factor system $\mu(g_i,g_j)=e^{-i h_i \cdot w_j}$, $h_i = R_i^{-1}k-k$.
Based on the factor system, we can construct the central extension $\bar{\mathcal{U}}_k^*$ of $\bar{\mathcal{U}}_k$, the group element in  $\bar{\mathcal{U}}_k^*$  takes the form of $(g,\alpha)$ $(g\in\bar{\mathcal{U}}_k,\,\alpha=1,2,\cdots,q-1)$, and the  group multiplication of
$\bar{\mathcal{U}}_k^*$ is defined as
\[
(g_i,\alpha)(g_i,\beta)=(g_ig_j,\,\alpha+\beta+a(g_i,g_j) \mod q),
\]
where the $a(g_i,g_j)$ and integer $q$ are determined by the factor system 
\[
\mu(g_i,g_j)=\exp{(2\pi i a(g_i,g_j)/q)}.
\]
Then the irreducible projective rep of $\bar{\mathcal{U}}_k$ is determined by
\[
\bar D_k^p(g_i)=\Delta_k^p((g_i,0)),
\] 
where $\Delta_k^p((g_i,\alpha))$ is the $p$-th allowed irreducible representation (irrep) of the central extension $\bar{\mathcal{U}}_k^*$ that satisfies
\[
\Delta_k^p((E,\alpha))=e^{2\pi i \alpha/q} I.
\]
Finally the small rep of ${\mathcal{U}}_k$ is given by
\[
D_k^p(\{S||R|t\})=\bar D_k^p(\{S||R\})e^{-ik\cdot t}.
\] 
Note that this method not only applies to the SLGs, but also to all the crystalline symmetry groups with translational symmetry.

In principle, such procedures can be applied to all types of SLGs. 
For non-collinear SLGs, the little co-group is always finite due to the finiteness of $\mathcal{S}_\mathrm{so}$, and its irreps can be effectively determined using the computational discrete algebra system \textsf{GAP} \cite{dabbaghian-abdoly_algorithm_2005, noauthor_gap_2024}. 
The resulting representation matrices have fully analytic forms and can be expressed as linear combinations of roots of unity (see Appendix~\ref{app:smallgroup}). 
However, for collinear SLGs, the little co-group at arbitrary $k$ point has a spin-only group $\mathcal{C}_\infty$ and is therefore infinite. 
As a result, the standard central-extension approach cannot be applied directly to such infinite groups. 
In the next subsection, we discuss how to construct the small coreps of collinear SLGs.

\subsection{Small coreps of collinear SLGs}
\label{sec:collinear}

For collinear SLGs, $\mathcal{U}_k$ has a semi-direct product structure of $\mathcal{U}'_k \ltimes \mathcal{C}_\infty$ [c.f. Eq. \eqref{eq:colUk}] in general.
Although the little co-group of $\mathcal{U}_k$ is infinite due to $\mathcal{C}_\infty$, we can first apply the central-extension approach to the little co-group of $\mathcal{U}'_k$ or its subgroup and then derive the small reps of $\mathcal{U}_k$.
The multiplication of any two elemenmts $u_1c_{\theta_1},u_2c_{\theta_2}\in\mathcal{U}_k$
 ($u_1,u_2\in \mathcal{U}_k^\prime $ and $c_{\theta_1},c_{\theta_2}\in \mathcal{C}_\infty$) can be written as
\[
u_1c_{\theta_1}u_2c_{\theta_2}=u_1u_2\varphi_{u_2}(c_{\theta_1})c_{\theta_2},
\]
where $\varphi$ is an homomorphism from $\mathcal{U}_k^\prime$ to a subgroup of $\text{Aut}(\mathcal{C}_\infty)$.
$\text{Aut}(\mathcal{C}_\infty)=\{\varepsilon,\rho\}$ is the automorphism group of $\mathcal{C}_{\infty}$, in which $\varepsilon$ is the identity map and $\rho$ reverses the rotation angle, i.e. 
\[
   \forall\theta:~~~\varepsilon(c_\theta)=c_\theta,~~~  \rho(c_\theta)=c_{-\theta}.
\]
The map $\varphi$ is determined via $\varphi_u$ as follows
\[
  \varphi_u(c_\theta)=u^{-1}c_\theta u~~~(u\in\mathcal{U}'_k,\,c_\theta\in\mathcal{C}_\infty).
\]
As mentioned in Sec. \ref{sec:SLGstruct}, the spin-rotation part of $\mathcal{U}'$ is $\{E\}$ for collinear SLGs.
This makes that $\varphi$ falls into two cases according to whether $\mathcal{U}'_k$ inherits this property.

\subsubsection{The case of $\mathcal{U}'_k=[\mathcal{U}']_k$}

In this case, $\mathcal{U}'_k\subseteq\mathcal{U}'$, and hence any $u\in\mathcal{U}'_k$  takes the form of $u=\{E||R|t\}$. This leads to $u^{-1}c_\theta u\equiv c_\theta$, i.e. $\varphi_u\equiv\varepsilon$, which is consistent with the direct product result
\[
\mathcal{U}_k=\mathcal{U}^\prime_k \times \mathcal C_\infty.
\]
Consequently, the small reps of $\mathcal{U}_k$ take a simple form
\[
D_{k,m}^p(uc_\theta)=D'^p_k(u)\Delta_m(c_\theta)~~~~(uc_\theta\in\mathcal{U}_k),
\]
where $D'^p_k$ is the $p$-th small rep of $\mathcal{U}_k^\prime$, and $\Delta_m(c_\theta) = e^{-i m \theta}$ $(m \in \mathbb{Z})$ is the irrep of $\mathcal{C}_\infty$. 
The small corep $\Gamma^p_{k,m}$ of $\mathcal{L}_k$ is just $D^p_{k,m}$ when $\mathcal{L}_k=\mathcal{U}_k$. In contrast, $\Gamma^p_{k,m}$ should be derived from $D^p_{k,m}$ according to the corep type $\tau^p_{k,m}$ if $\mathcal{L}_k=\mathcal{U}_k+\mathcal{AU}_k$.

According to Eq.~\eqref{eq:Ukp}, FM falls into the case of $\mathcal{U}'_k=[\mathcal{U}']_k$. Besides, AFM or AM also fall into this case when $[U'_0\mathcal{U}']_k=\varnothing$.  
Denote the four terms of $\mathcal{L}_k$ in Eq. \eqref{eq:AFMLk} as $\Lambda_i\mathcal{C}_\infty$  $(i=1,2,3,4)$ respectively. 
Accordingly, 
\[
\mathcal{U}'_k=[\mathcal{U}']_k=\Lambda_1,
\]
and $\mathcal{L}_k=\mathcal{U}_k+\mathcal{AU}_k$ takes two forms
\[
\mathcal{L}_k=
\begin{cases}
   \Lambda_1\mathcal{C}_\infty+\Lambda_3\mathcal{C}_\infty = \mathcal{U}_k+\mathcal{A}_1\mathcal{U}_k  &  \text{(F1)}, \\
   \Lambda_1\mathcal{C}_\infty+\Lambda_4\mathcal{C}_\infty = \mathcal{U}_k+\mathcal{A}_2\mathcal{U}_k  &  \text{(F2)},
\end{cases}
\]
where $\mathcal{A}_1=\{C_{2x}\mathcal{T}||U_3\}$ and $\mathcal{A}_2=\{\mathcal{T}||U_4\}$.  
Here, $U_3\in[\mathcal{U}']_{-k}$ and $U_4\in[U'_0\mathcal{U}']_{-k}$ are both unitary spatial operation.
FM only takes the (F1) form, while AFM or AM may take both (F1) and (F2) forms.
These two forms of $\mathcal{L}_k$ make corep type $\tau^p_{k,m}$ behave differently.
Taking the form $g=\mathcal{A}uc_\theta$ for $\forall g\in\mathcal{AU}_k$, we have
\[
\begin{split}
	\tau_{k,m}^{p}= & \frac{1}{|\mathcal{U}_{k}|}\sum_{g\in \mathcal{A}\mathcal{U}_{k}}\chi_{k,m}^{p}(g^{2})\\
	              = & \frac{1}{|\mathcal{U}'_{k}|}\frac{1}{2\pi}\sum_{u\in [\mathcal{U}']_{k}} \int_0^{2\pi} \chi_{k,m}^{p}((\mathcal{A}uc_\theta)^{2}) \,\mathrm{d}\theta \\
	= & \frac{1}{|\mathcal{U}'_{k}|}\sum_{u\in [\mathcal{U}']_{k}}\chi_{k}^{\prime p}((\mathcal{A}u)^{2}) \underbrace{\frac1{2\pi} \int_0^{2\pi}\Delta_{m}(c_{\theta'}c_\theta)\,\mathrm{d}\theta}_{\gamma}
\end{split}
\]
where $\chi^p_{k,m}$ and $\chi'^p_k$ are the characters of $D^p_{k,m}$ and $D'^p_k$ respectively.
We have used the relation $(\mathcal{A}uc_\theta)^2=(Au)^2c_{\theta'}c_\theta$ 
where $c_{\theta'}=\mathcal{A}^{-1}c_\theta\mathcal{A}$ behaves differently for (F1) and (F2)
\[
  c_{\theta'}=\begin{cases}
     	\mathcal{A}_1^{-1}c_\theta\mathcal{A}_1 = c_{-\theta} & \text{(F1)}, \\
     	\mathcal{A}_2^{-1}c_\theta\mathcal{A}_2 = c_{\theta} & \text{(F2)} .\\
  \end{cases}
\]
Consequently, we have
\[
   \gamma=\begin{cases}
   \frac1{2\pi}\int_0^{2\pi} \Delta_m(E)\,\mathrm{d}\theta = 1 & \text{(F1)}, \\
   \frac1{2\pi}\int_0^{2\pi} \Delta_m(c_{2\theta})\,\mathrm{d}\theta = \begin{cases}
   1 & m=0 \\
   0 & m\ne 0
   \end{cases} & \text{(F2)}.
   \end{cases}
\]
Letting
\[
  \sigma_1=\sum_{a\in\Lambda_3} \chi'^p_k(a^2) \text{~~~and~~~}
  \sigma_2=\sum_{a\in\Lambda_4} \chi'^p_k(a^2),
\]
the corep type is Finally
\[
   \tau^p_{k,m}=\begin{cases}
	 \sigma_1 / |\mathcal{U}'_k|  & \text{(F1)}, \\
	 \sigma_2 / |\mathcal{U}'_k|  & \text{(F2) and }m= 0, \\
	 0  & \text{(F2) and }m\ne 0. 
   \end{cases}
\]


\subsubsection{The case of $\mathcal{U}'_k\supset[\mathcal{U}']_k$}

In this case, all four terms in Eq. \eqref{eq:AFMLk} are nonempty, and we have
\[
    \mathcal{U}'_k=\Lambda_1+\Lambda_2=[\mathcal{U}']_k + r[\mathcal{U}']_k
\]
with $r=\{C_{2x}||U_2\}$ and $U_2\in[U'_0\mathcal{U}']_k$. 
Since $r^{-1}c_\theta r=c_{-\theta}=\rho(c_\theta)$, the homomorphism $\varphi$ is now
\[
    \varphi_u=\begin{cases}
		\varepsilon  &  u\in[\mathcal{U}']_k, \\
		\rho&  u\in r[\mathcal{U}']_k.
	\end{cases}
\]
This case occurs only in AFM or AM systems.

We now apply the general little group theory to derive the  irreps of $\mathcal{U}_k$. 
Since $\mathcal{C}_\infty$ is a normal subgroup of $\mathcal{U}_k$ and its irreps are
$\Delta_m(c_\theta)=e^{- i m \theta }$,  each $g\in\mathcal{U}_k$ maps the irrep $\Delta_m$ to a conjugate rep defined by
\[
\Delta_m^g(c_\theta) \equiv \Delta_m(g^{-1}c_\theta g)=\begin{cases}
	\Delta_m(c_\theta )  & g\in \mathcal {V}\\
	\Delta_{-m}(c_\theta ) & g\in  r\mathcal {V}.\\
\end{cases}
\]
where $\mathcal{V}=[\mathcal{U}']_k\mathcal{C}_\infty$ and hence $\mathcal{U}_k=\mathcal{V}+r\mathcal{V}$.
Then the little group (stabilizer) for $\Delta_m$ is 
\[
   \mathcal{U}_k^m=\{ g\in\mathcal{U}_k \mid \Delta_m^g=\Delta_m\}.
\]
Depending on the value of $m$, the resulting irreps fall into two categories:

\textbf{Case i}: $m=0$. The irrep $\Delta_0$ is invariant under any $g = uc_\theta\in \mathcal{U}_k$ ($u\in\mathcal{U}'_k$). Thus, $\mathcal{U}_k^m=\mathcal{U}_k$, and the small rep of $\mathcal{U}_k$ are simply 
\[
D_{k,m}^p(uc_\theta)=D_k^{\prime p}(u), \\
\]
where $D_k^{\prime p}$ is the small rep of $\mathcal{U}_k^\prime$.

\textbf{Case ii}, $m \neq 0$. The orbit of $\Delta_m$ contains two irreps $\{\Delta_m, \Delta_{-m}\}$. The little group $\mathcal{U}_k^m$ equals $\mathcal{V}$. The small rep of $\mathcal{U}_k$ is induced from the small rep of $\mathcal{V}$. For $g \in \mathcal{V}$, the rep is diagonal
\[
D_{k,m}^p(uc_\theta)= \begin{pmatrix} \Delta_m(c_\theta)  F_k^{\prime p}(u)& 0 \\ 0 & \Delta_{-m}(c_\theta)  F_k^{\prime p}(r^{-1}ur) \end{pmatrix},
\]
while for $g \in r\mathcal{V}$ it takes the off-diagonal form
\[
D_{k,m}^p(uc_\theta)= \begin{pmatrix} 0& \Delta_{-m}(c_\theta)  F_k^{\prime p}(ur) \\  \Delta_{m}(c_\theta)  F_k^{\prime p}(r^{-1}u)
& 0
 \end{pmatrix}.
\]
where $F_k^{\prime p}$ is the small rep of $[\mathcal{U}']_k$.

To calculate the corep type $\frac{1}{|\mathcal{U}_{k}|}\sum_{g\in \mathcal{A}\mathcal{U}_{k}}\chi_{k,m}^{p}(g^{2})$, the range of summation is now  $\mathcal{AU}_k=\Lambda_3\mathcal{C}_\infty+\Lambda_4\mathcal{C}_\infty$.  
This range is the union of the ranges in cases (F1) and (F2). The integral over $\theta$ behaves in the same way as $\gamma$ under the correspondence
\[
  \text{(F1)}~\Leftrightarrow~\mathcal{A}u\in\Lambda_3,~~~
  \text{(F2)}~\Leftrightarrow~\mathcal{A}u\in\Lambda_4.
\]
Consequently, the corep type of $D^p_{k,m}$ is
\[
    \tau^p_{k,m}=\begin{cases}
      \frac{1}{|\mathcal{U}'_k|} \sum\limits_{a\in (\Lambda_3+\Lambda_4)} [\psi'^p_k(a^2)+\psi'^p_k(r^{-1}a^2r)]  & m=0 \\
      \frac{1}{|\mathcal{U}'_k|} \sum\limits_{a\in \Lambda_3} [\psi'^p_k(a^2)+\psi'^p_k(r^{-1}a^2r)]  & m\ne 0
    \end{cases}
\]
where $\psi'^p_k$ is the character of $F'^p_k$.



\begin{table}[t]
	\caption{Number of SLGs ($I_k\leq8$) that host different types of degeneracies.
	HSP, HSL, and GP means high-symmetry point, high-symmetry line, and general point, respectively.
	$k$ points can be classified into  HSP, HSL, and GP using the method in Appendix~\ref{app:findk}. 
	``$n$ (HSP/HSL/GP)'' means $n$-fold degeneracy, corresponding to $n$-dimensional small corep, at/on/at HSP/HSL/GP.
	}
	\label{tab:slgsum}
	\begin{ruledtabular}
		\begin{tabular}{llll}
			Degeneracy type&Collinear& Coplanar & Non-coplanar \\
			\hline
			\textbf{Single SLGs}  	&&&  \\
			1 (HSP) &  448 &4306& 17537\\
			2 (HSP) & 429 	&4154	&17150\\
			3 (HSP) &0 		&31		&168\\
			4 (HSP) &242	&1702	&6745\\
			8 (HSP) &11	&0		&0\\			
			1 (HSL) &  395 &4051&21983\\
			2 (HSL) &363		&3612 	&18832\\
			4 (HSL) &141	&0		&0\\
			1 (GP) &448	&	4306& 22775\\
			2 (GP)  &276		&0 		&0\\
			&&&  \\
			\textbf{Double SLGs}		&&&   \\
			1 (HSP) &  95 &1912& 7748\\
			2 (HSP) &429 	&4247	&17342\\
			3 (HSP) &0 		&18		&83\\
			4 (HSP) &242	&2308	&8613\\
			6 (HSP) &0		&13		&34\\
			8 (HSP) &11	&60		&114\\			
			1 (HSL) &  88 &2964& 17104\\
			2 (HSL) &363		&3904 	&20074\\
			4 (HSL) &141	&467	&1177\\
			1 (GP) &172	&3272	&19327\\
			2 (GP)  &276		&1034	&3448\\
		\end{tabular}
	\end{ruledtabular}
\end{table}

\subsection{Single-valued and double-valued coreps of SLGs}
\label{sec:single}

The symmetry analysis of electronic systems with spin-$1/2$ degrees of freedom requires the use of the double groups of SLGs, namely the double covers of the corresponding single SLGs. While the single SLGs are sufficient for bosonic or artificial systems, the double groups are indispensable for describing electronic band structures.

To derive the coreps of double SLGs, one need also first determine the double-valued irreps of the double SLGs and then calculate the corep types of the irreps. The key distinction for double-valued irreps is that the character of the $2\pi$ rotation (denoted as $\bar{E}$) satisfies the condition $\chi(\bar{E}) = -\chi(E)$. Furthermore, the time-reversal operator must satisfy $\mathcal{T}^2 = \bar E$ for fermionic systems, in contrast to $\mathcal{T}^2 = E$ in the single-valued case. In coplanar and non-coplanar SLGs, which consist of discrete symmetry operations, these double-valued coreps can be obtained through standard procedure as described in Sec. \ref{subsec:corep}-\ref{subsec:rep}.

However, the collinear case requires a more careful treatment of the continuous group $\mathcal{C}_\infty$, which is extended to its double cover $\bar{\mathcal{C}}_\infty$.   The rotation angle $\theta$ for the element $c_{\theta} \in \bar{\mathcal{C}}_\infty$ spans the range $[0, 4\pi)$. The isomorphism between ${\mathcal{C}}_\infty$ and $\bar{ \mathcal{C}}_\infty$ (defined via the bijection $c_\theta \in \mathcal{C}_\infty \leftrightarrow c_{2\theta} \in \bar{\mathcal{C}}_\infty$) ensures that their automorphism groups are the same, and the irreps of the $\bar{\mathcal{C}}_\infty$ take the general form $\Delta_n(c_\theta) = e^{-i \frac{n}{2}\theta}$ ($n \in \mathbb{Z}$). To obtain the double-valued irreps of $\bar{\mathcal{C}}_\infty$, one selects the irreps satisfying $\chi(\bar{E})=\chi(c_{2\pi}) = -\chi(E)=-1$. By setting $n = 2m+1$ ($m \in \mathbb{Z}$), these double-valued irreps are expressed as $\bar{\Delta}_m(c_\theta) = e^{-i (m+\frac{1}{2}) \theta}$. Accordingly,  for the evaluation of coreps types, the integration is adjusted from$ \frac1{2\pi}\int_0^{2\pi} \cdots \,\mathrm{d}\theta$ to $\frac1{4\pi}\int_0^{4\pi} \cdots \mathrm{d}\theta$.

\section{Summary}

The type of symmetry-enforced degeneracies in the 27,529 SLGs with $I_k \leq 8$ are summarized in Table~\ref{tab:slgsum}. 
Compared with magnetic layer groups, in which spin and spatial rotations are locked, SLGs exhibit a significantly larger variety of possible band degeneracies, particularly high-fold degeneracies. 
The underlying group structure plays a central role in determining the degeneracy types. 
A comprehensive quantitative mapping between all degeneracy types and the corresponding SLGs is provided in S4 of the Supplemental Material \cite{noauthor_notitle_nodate}. {\color{black}
	
It should be mentioned  that
the single-valued coreps of collinear SLGs behave fundamentally differently from those of standard space groups. 	 
In conventional single space groups, the combination of spatial inversion $P$ and time reversal symmetry
$\mathcal{T}$  does not enforce additional band degeneracy, while double space groups always enforce at least double degeneracy due to $(P\mathcal{T})^2=\bar E$. 
However, A single collinear SLG with 
$\{\mathcal{T}||P|\boldsymbol{0}\}$ 
symmetry can indeed enforce double degeneracy. For example, consider a  $\boldsymbol{k}$ point whose single (double) little co-group is
$
(\{E||E\}+\{\mathcal{T}||P\})\times \mathcal{C}_\infty (\bar{\mathcal{C}}_\infty)
$, which is the typical case for a generic $\boldsymbol{k}$ point.
Then the corep types at $\boldsymbol{k}$ are:
\[
\begin{split}
\tau_{k,m}&=\begin{cases}\frac1{2\pi} \int_0^{2\pi}e^{-2 i m \theta}\,\mathrm{d}\theta& \text{Single SLGs}\\
	-\frac1{4\pi} \int_0^{4\pi}e^{-2 i (m+\frac{1}{2})\theta}\,\mathrm{d}\theta& \text{Double SLGs}\\
	\end{cases}\\
&=\begin{cases}
	1 & m=0,  \text{single SLGs}\\
	0 & m\ne 0, \text{single SLGs}\\
		0 & m\in \mathbb{Z}, \text{double SLGs}.
\end{cases}
\end{split}
\]
Specifically, $\tau=1$ corresponds to a case-($a$) corep (non-degenerate), while $\tau=0$ indicates a case-($c$) corep, where two irreps pair up to form a doubly degenerate band. Consequently, for double SLGs with $P\mathcal{T}$ symmetry, the band structure is at least doubly degenerate for all $m$. In contrast, for single SLGs, the $P\mathcal{T}$ symmetry allows for the coexistence of both non-degenerate ($m=0$) and doubly degenerate ($m \neq 0$) states.

}

In conclusion, we systematically determine the group structures of 40,087 SLGs and analytically derive the small coreps of 27,529 SLGs. 
This work establishes a comprehensive theoretical framework for the symmetry analysis of 2D magnetic systems and provides a practical foundation for the discovery and application of novel 2D magnetic materials. 
Our results can be used to search for emergent particles in low-dimensional systems and to facilitate the symmetry-guided design of low-dimensional magnetic crystals.

\section*{Supplemental Material}
The Supplemental Material, which includes complete classification results and exhaustive mapping tables for all 40,087 SLGs, is available as an ancillary file on arXiv. To download it, click “TeX Source” or “Ancillary files” in the right panel of the arXiv article page.
\begin{acknowledgments}

This work is supported by the National Natural Science Foundation of China (Grant Nos. 12274028, 12234003 and 12274028) and the National Key R\&D Program of China (2022YFA1402603).

\end{acknowledgments}




\appendix
\section{Solving the linear congruence equations module 1}
\label{app:lce}
Consider the following congruence equations:
\[
M\mathbf{x} = \mathbf{b} \mod 1
\]
where \( M \) is an \( m \times n \) integer matrix, and \( \mathbf{b} \) is a column vector whose elements may be fractions (fractional translations). Perform the Smith decomposition on matrix \( M \) \cite{cohen_course_1993, eick_computing_1997}:
\[
UMV = D
\]
where \( D \) is an \( m \times n \) integer matrix with only non-zero elements on the diagonal, and \( U \) and \( V \) are \( m \times m \) and \( n \times n \) integer matrices, respectively, whose inverses are also integer matrices, and \( |\det U| = |\det V| = 1 \). First, solve
\[
D\mathbf{x}' = U\mathbf{b} \mod 1
\]
where \( \mathbf{x}' = V^{-1}\mathbf{x} \). Since \( D \) is a diagonal matrix, this equation essentially reduces to a series of single-variable linear congruence equations. By multiplying both sides of the equation by the least common multiple of the denominators of \( U\mathbf{b} \), the Chinese Remainder Theorem can be applied to solve it. Finally, all possible solutions to the original equation are given by:
\[
\mathbf{x} = V\mathbf{x}' \mod 1.
\]

\section{Finding high-symmetry \texorpdfstring{$k$}{k} points in SLGs}
\label{app:findk}

To identify all high-symmetry $k$ points of a given SLG $\mathcal{L}$, we first extract the spin point group $P$ of the nontrivial part $\mathcal{L}'$ of $\mathcal{L}$, 
\[
  P=\{ \{S||R\} \mid \{S||R|t\}\in\mathcal{L}' \}.
\]
We then enumerate all subgroups of $P$.
For each subgroup, let $\{R_i\}$ denote the set of spatial-rotation parts of its elements. The $k$ points invariant under this subgroup are obtained by solving the following system of linear congruence equations:
\[
R_i k=
\begin{cases}
k+K_m, & \text{for unitary elements},\\
-k+K_m, & \text{for anti-unitary elements},
\end{cases}
\]
where $K_m$ is a reciprocal lattice vector.
Repeating this procedure for every subgroup of $P$ generates all symmetry-distinct high-symmetry $k$ points of the SLG. 

Since $P$ is finite groups, all its subgroups can be systematically enumerated using \textsf{GAP}. 
The linear congruence equations are also solved using the method described in Appendix~\ref{app:lce}.

\section{Irreducible representations of abstract groups}
\label{app:smallgroup}
In all cases considered in this work, the central extension of the little co-group is finite.
Therefore, we can always find a finite abstract group isomorphic to the central extension.
For groups of order less than 2000, excluding those of orders 512 and 1024, we directly use the Small Groups Library in \textsf{GAP}, which contains the irreps required in this work \cite{noauthor_gap_2024}. 
For the remaining cases, including groups of orders 512, 1024, and 2048, we first construct an isomorphic polycyclic group in \textsf{GAP} and then compute its irreps using Dixon's algorithm for polycyclic groups \cite{dabbaghian-abdoly_algorithm_2005}.

\bibliography{SLG}

\begin{thebibliography}{35}%
\makeatletter
\providecommand \@ifxundefined [1]{%
 \@ifx{#1\undefined}
}%
\providecommand \@ifnum [1]{%
 \ifnum #1\expandafter \@firstoftwo
 \else \expandafter \@secondoftwo
 \fi
}%
\providecommand \@ifx [1]{%
 \ifx #1\expandafter \@firstoftwo
 \else \expandafter \@secondoftwo
 \fi
}%
\providecommand \natexlab [1]{#1}%
\providecommand \enquote  [1]{``#1''}%
\providecommand \bibnamefont  [1]{#1}%
\providecommand \bibfnamefont [1]{#1}%
\providecommand \citenamefont [1]{#1}%
\providecommand \href@noop [0]{\@secondoftwo}%
\providecommand \href [0]{\begingroup \@sanitize@url \@href}%
\providecommand \@href[1]{\@@startlink{#1}\@@href}%
\providecommand \@@href[1]{\endgroup#1\@@endlink}%
\providecommand \@sanitize@url [0]{\catcode `\\12\catcode `\$12\catcode
  `\&12\catcode `\#12\catcode `\^12\catcode `\_12\catcode `\%12\relax}%
\providecommand \@@startlink[1]{}%
\providecommand \@@endlink[0]{}%
\providecommand \url  [0]{\begingroup\@sanitize@url \@url }%
\providecommand \@url [1]{\endgroup\@href {#1}{\urlprefix }}%
\providecommand \urlprefix  [0]{URL }%
\providecommand \Eprint [0]{\href }%
\providecommand \doibase [0]{https://doi.org/}%
\providecommand \selectlanguage [0]{\@gobble}%
\providecommand \bibinfo  [0]{\@secondoftwo}%
\providecommand \bibfield  [0]{\@secondoftwo}%
\providecommand \translation [1]{[#1]}%
\providecommand \BibitemOpen [0]{}%
\providecommand \bibitemStop [0]{}%
\providecommand \bibitemNoStop [0]{.\EOS\space}%
\providecommand \EOS [0]{\spacefactor3000\relax}%
\providecommand \BibitemShut  [1]{\csname bibitem#1\endcsname}%
\let\auto@bib@innerbib\@empty
\bibitem [{\citenamefont {Dresselhaus}\ \emph {et~al.}(2008)\citenamefont
  {Dresselhaus}, \citenamefont {Dresselhaus},\ and\ \citenamefont
  {Jorio}}]{dresselhaus_group_2008}%
  \BibitemOpen
  \bibfield  {author} {\bibinfo {author} {\bibfnamefont {M.~S.}\ \bibnamefont
  {Dresselhaus}}, \bibinfo {author} {\bibfnamefont {G.}~\bibnamefont
  {Dresselhaus}},\ and\ \bibinfo {author} {\bibfnamefont {A.}~\bibnamefont
  {Jorio}},\ }\href@noop {} {\emph {\bibinfo {title} {Group theory: application
  to the physics of condensed matter}}}\ (\bibinfo  {publisher}
  {Springer-Verlag},\ \bibinfo {address} {Berlin},\ \bibinfo {year}
  {2008})\BibitemShut {NoStop}%
\bibitem [{\citenamefont {Bradley}\ and\ \citenamefont
  {Cracknell}(2009)}]{bradley_mathematical_2009}%
  \BibitemOpen
  \bibfield  {author} {\bibinfo {author} {\bibfnamefont {C.}~\bibnamefont
  {Bradley}}\ and\ \bibinfo {author} {\bibfnamefont {A.}~\bibnamefont
  {Cracknell}},\ }\href@noop {} {\emph {\bibinfo {title} {Mathematical theory
  of symmetry in solids: representation theory for point groups and space
  groups}}}\ (\bibinfo  {publisher} {OUP},\ \bibinfo {address} {Oxford},\
  \bibinfo {year} {2009})\BibitemShut {NoStop}%
\bibitem [{\citenamefont {Litvin}(2016)}]{litvin_magnetic_2016}%
  \BibitemOpen
  \bibfield  {author} {\bibinfo {author} {\bibfnamefont {D.~B.}\ \bibnamefont
  {Litvin}},\ }\href {https://doi.org/10.1107/9780955360220001} {\emph
  {\bibinfo {title} {Magnetic {Subperiodic} {Groups} and {Magnetic} {Space}
  {Groups}}}}\ (\bibinfo  {publisher} {Wiley Online Library},\ \bibinfo {year}
  {2016})\BibitemShut {NoStop}%
\bibitem [{\citenamefont {Brinkman}\ and\ \citenamefont
  {Elliott}(1966)}]{w_f_brinkman_theory_1966}%
  \BibitemOpen
  \bibfield  {author} {\bibinfo {author} {\bibfnamefont {W.~F.}\ \bibnamefont
  {Brinkman}}\ and\ \bibinfo {author} {\bibfnamefont {R.~J.}\ \bibnamefont
  {Elliott}},\ }\bibfield  {title} {\bibinfo {title} {Theory of spin-space
  groups},\ }\href {https://doi.org/10.1098/rspa.1966.0211} {\bibfield
  {journal} {\bibinfo  {journal} {Proceedings of the Royal Society of London.
  Series A. Mathematical and Physical Sciences}\ }\textbf {\bibinfo {volume}
  {294}},\ \bibinfo {pages} {343} (\bibinfo {year} {1966})}\BibitemShut
  {NoStop}%
\bibitem [{\citenamefont {Litvin}\ and\ \citenamefont
  {Opechowski}(1974)}]{litvin_spin_1974}%
  \BibitemOpen
  \bibfield  {author} {\bibinfo {author} {\bibfnamefont {D.}~\bibnamefont
  {Litvin}}\ and\ \bibinfo {author} {\bibfnamefont {W.}~\bibnamefont
  {Opechowski}},\ }\bibfield  {title} {\bibinfo {title} {Spin {Groups}},\
  }\href@noop {} {\bibfield  {journal} {\bibinfo  {journal} {Physica}\ }\textbf
  {\bibinfo {volume} {76}},\ \bibinfo {pages} {538} (\bibinfo {year}
  {1974})}\BibitemShut {NoStop}%
\bibitem [{\citenamefont {Litvin}(1977)}]{litvin_spin_1977}%
  \BibitemOpen
  \bibfield  {author} {\bibinfo {author} {\bibfnamefont {D.~B.}\ \bibnamefont
  {Litvin}},\ }\bibfield  {title} {\bibinfo {title} {Spin point groups},\
  }\href {https://doi.org/10.1107/S0567739477000709} {\bibfield  {journal}
  {\bibinfo  {journal} {Acta Crystallographica Section A}\ }\textbf {\bibinfo
  {volume} {33}},\ \bibinfo {pages} {279} (\bibinfo {year} {1977})}\BibitemShut
  {NoStop}%
\bibitem [{\citenamefont {Chen}\ \emph {et~al.}(2024)\citenamefont {Chen},
  \citenamefont {Ren}, \citenamefont {Zhu}, \citenamefont {Yu}, \citenamefont
  {Zhang}, \citenamefont {Liu}, \citenamefont {Li}, \citenamefont {Liu},
  \citenamefont {Li},\ and\ \citenamefont {Liu}}]{chen_enumeration_2024}%
  \BibitemOpen
  \bibfield  {author} {\bibinfo {author} {\bibfnamefont {X.}~\bibnamefont
  {Chen}}, \bibinfo {author} {\bibfnamefont {J.}~\bibnamefont {Ren}}, \bibinfo
  {author} {\bibfnamefont {Y.}~\bibnamefont {Zhu}}, \bibinfo {author}
  {\bibfnamefont {Y.}~\bibnamefont {Yu}}, \bibinfo {author} {\bibfnamefont
  {A.}~\bibnamefont {Zhang}}, \bibinfo {author} {\bibfnamefont
  {P.}~\bibnamefont {Liu}}, \bibinfo {author} {\bibfnamefont {J.}~\bibnamefont
  {Li}}, \bibinfo {author} {\bibfnamefont {Y.}~\bibnamefont {Liu}}, \bibinfo
  {author} {\bibfnamefont {C.}~\bibnamefont {Li}},\ and\ \bibinfo {author}
  {\bibfnamefont {Q.}~\bibnamefont {Liu}},\ }\bibfield  {title} {\bibinfo
  {title} {Enumeration and {Representation} {Theory} of {Spin} {Space}
  {Groups}},\ }\href {https://doi.org/10.1103/PhysRevX.14.031038} {\bibfield
  {journal} {\bibinfo  {journal} {Physical Review X}\ }\textbf {\bibinfo
  {volume} {14}},\ \bibinfo {pages} {031038} (\bibinfo {year}
  {2024})}\BibitemShut {NoStop}%
\bibitem [{\citenamefont {Jiang}\ \emph {et~al.}(2024)\citenamefont {Jiang},
  \citenamefont {Song}, \citenamefont {Zhu}, \citenamefont {Fang},
  \citenamefont {Weng}, \citenamefont {Liu}, \citenamefont {Yang},\ and\
  \citenamefont {Fang}}]{jiang_enumeration_2024}%
  \BibitemOpen
  \bibfield  {author} {\bibinfo {author} {\bibfnamefont {Y.}~\bibnamefont
  {Jiang}}, \bibinfo {author} {\bibfnamefont {Z.}~\bibnamefont {Song}},
  \bibinfo {author} {\bibfnamefont {T.}~\bibnamefont {Zhu}}, \bibinfo {author}
  {\bibfnamefont {Z.}~\bibnamefont {Fang}}, \bibinfo {author} {\bibfnamefont
  {H.}~\bibnamefont {Weng}}, \bibinfo {author} {\bibfnamefont {Z.-X.}\
  \bibnamefont {Liu}}, \bibinfo {author} {\bibfnamefont {J.}~\bibnamefont
  {Yang}},\ and\ \bibinfo {author} {\bibfnamefont {C.}~\bibnamefont {Fang}},\
  }\bibfield  {title} {\bibinfo {title} {Enumeration of {Spin}-{Space}
  {Groups}: {Toward} a {Complete} {Description} of {Symmetries} of {Magnetic}
  {Orders}},\ }\href {https://doi.org/10.1103/PhysRevX.14.031039} {\bibfield
  {journal} {\bibinfo  {journal} {Physical Review X}\ }\textbf {\bibinfo
  {volume} {14}},\ \bibinfo {pages} {031039} (\bibinfo {year}
  {2024})}\BibitemShut {NoStop}%
\bibitem [{\citenamefont {Xiao}\ \emph {et~al.}(2024)\citenamefont {Xiao},
  \citenamefont {Zhao}, \citenamefont {Li}, \citenamefont {Shindou},\ and\
  \citenamefont {Song}}]{xiao_spin_2024}%
  \BibitemOpen
  \bibfield  {author} {\bibinfo {author} {\bibfnamefont {Z.}~\bibnamefont
  {Xiao}}, \bibinfo {author} {\bibfnamefont {J.}~\bibnamefont {Zhao}}, \bibinfo
  {author} {\bibfnamefont {Y.}~\bibnamefont {Li}}, \bibinfo {author}
  {\bibfnamefont {R.}~\bibnamefont {Shindou}},\ and\ \bibinfo {author}
  {\bibfnamefont {Z.-D.}\ \bibnamefont {Song}},\ }\bibfield  {title} {\bibinfo
  {title} {Spin {Space} {Groups}: {Full} {Classification} and {Applications}},\
  }\href {https://doi.org/10.1103/PhysRevX.14.031037} {\bibfield  {journal}
  {\bibinfo  {journal} {Physical Review X}\ }\textbf {\bibinfo {volume} {14}},\
  \bibinfo {pages} {031037} (\bibinfo {year} {2024})}\BibitemShut {NoStop}%
\bibitem [{\citenamefont {Liu}\ \emph {et~al.}(2022)\citenamefont {Liu},
  \citenamefont {Li}, \citenamefont {Han}, \citenamefont {Wan},\ and\
  \citenamefont {Liu}}]{liu_spin-group_2022}%
  \BibitemOpen
  \bibfield  {author} {\bibinfo {author} {\bibfnamefont {P.}~\bibnamefont
  {Liu}}, \bibinfo {author} {\bibfnamefont {J.}~\bibnamefont {Li}}, \bibinfo
  {author} {\bibfnamefont {J.}~\bibnamefont {Han}}, \bibinfo {author}
  {\bibfnamefont {X.}~\bibnamefont {Wan}},\ and\ \bibinfo {author}
  {\bibfnamefont {Q.}~\bibnamefont {Liu}},\ }\bibfield  {title} {\bibinfo
  {title} {Spin-{Group} {Symmetry} in {Magnetic} {Materials} with {Negligible}
  {Spin}-{Orbit} {Coupling}},\ }\href
  {https://doi.org/10.1103/PhysRevX.12.021016} {\bibfield  {journal} {\bibinfo
  {journal} {Physical Review X}\ }\textbf {\bibinfo {volume} {12}},\ \bibinfo
  {pages} {021016} (\bibinfo {year} {2022})}\BibitemShut {NoStop}%
\bibitem [{\citenamefont {Chen}\ \emph {et~al.}(2025)\citenamefont {Chen},
  \citenamefont {Liu}, \citenamefont {Liu}, \citenamefont {Yu}, \citenamefont
  {Ren}, \citenamefont {Li}, \citenamefont {Zhang},\ and\ \citenamefont
  {Liu}}]{chen_unconventional_2025}%
  \BibitemOpen
  \bibfield  {author} {\bibinfo {author} {\bibfnamefont {X.}~\bibnamefont
  {Chen}}, \bibinfo {author} {\bibfnamefont {Y.}~\bibnamefont {Liu}}, \bibinfo
  {author} {\bibfnamefont {P.}~\bibnamefont {Liu}}, \bibinfo {author}
  {\bibfnamefont {Y.}~\bibnamefont {Yu}}, \bibinfo {author} {\bibfnamefont
  {J.}~\bibnamefont {Ren}}, \bibinfo {author} {\bibfnamefont {J.}~\bibnamefont
  {Li}}, \bibinfo {author} {\bibfnamefont {A.}~\bibnamefont {Zhang}},\ and\
  \bibinfo {author} {\bibfnamefont {Q.}~\bibnamefont {Liu}},\ }\bibfield
  {title} {\bibinfo {title} {Unconventional magnons in collinear magnets
  dictated by spin space groups},\ }\bibfield  {journal} {\bibinfo  {journal}
  {Nature}\ }\href {https://doi.org/10.1038/s41586-025-08715-7}
  {10.1038/s41586-025-08715-7} (\bibinfo {year} {2025})\BibitemShut {NoStop}%
\bibitem [{\citenamefont {Šmejkal}\ \emph {et~al.}(2022)\citenamefont
  {Šmejkal}, \citenamefont {Sinova},\ and\ \citenamefont
  {Jungwirth}}]{smejkal_beyond_2022}%
  \BibitemOpen
  \bibfield  {author} {\bibinfo {author} {\bibfnamefont {L.}~\bibnamefont
  {Šmejkal}}, \bibinfo {author} {\bibfnamefont {J.}~\bibnamefont {Sinova}},\
  and\ \bibinfo {author} {\bibfnamefont {T.}~\bibnamefont {Jungwirth}},\
  }\bibfield  {title} {\bibinfo {title} {Beyond {Conventional} {Ferromagnetism}
  and {Antiferromagnetism}: {A} {Phase} with {Nonrelativistic} {Spin} and
  {Crystal} {Rotation} {Symmetry}},\ }\href
  {https://doi.org/10.1103/PhysRevX.12.031042} {\bibfield  {journal} {\bibinfo
  {journal} {Physical Review X}\ }\textbf {\bibinfo {volume} {12}},\ \bibinfo
  {pages} {031042} (\bibinfo {year} {2022})}\BibitemShut {NoStop}%
\bibitem [{\citenamefont {Jungwirth}\ \emph {et~al.}(2025)\citenamefont
  {Jungwirth}, \citenamefont {Fernandes}, \citenamefont {Fradkin},
  \citenamefont {MacDonald}, \citenamefont {Sinova},\ and\ \citenamefont
  {Šmejkal}}]{jungwirth_altermagnetism_2025}%
  \BibitemOpen
  \bibfield  {author} {\bibinfo {author} {\bibfnamefont {T.}~\bibnamefont
  {Jungwirth}}, \bibinfo {author} {\bibfnamefont {R.~M.}\ \bibnamefont
  {Fernandes}}, \bibinfo {author} {\bibfnamefont {E.}~\bibnamefont {Fradkin}},
  \bibinfo {author} {\bibfnamefont {A.~H.}\ \bibnamefont {MacDonald}}, \bibinfo
  {author} {\bibfnamefont {J.}~\bibnamefont {Sinova}},\ and\ \bibinfo {author}
  {\bibfnamefont {L.}~\bibnamefont {Šmejkal}},\ }\bibfield  {title} {\bibinfo
  {title} {Altermagnetism: {An} unconventional spin-ordered phase of matter},\
  }\href {https://doi.org/10.1016/j.newton.2025.100162} {\bibfield  {journal}
  {\bibinfo  {journal} {Newton}\ }\textbf {\bibinfo {volume} {1}},\ \bibinfo
  {pages} {100162} (\bibinfo {year} {2025})}\BibitemShut {NoStop}%
\bibitem [{\citenamefont {Huang}\ \emph {et~al.}(2017)\citenamefont {Huang},
  \citenamefont {Clark}, \citenamefont {Navarro-Moratalla}, \citenamefont
  {Klein}, \citenamefont {Cheng}, \citenamefont {Seyler}, \citenamefont
  {Zhong}, \citenamefont {Schmidgall}, \citenamefont {McGuire}, \citenamefont
  {Cobden}, \citenamefont {Yao}, \citenamefont {Xiao}, \citenamefont
  {Jarillo-Herrero},\ and\ \citenamefont {Xu}}]{huang_layer-dependent_2017}%
  \BibitemOpen
  \bibfield  {author} {\bibinfo {author} {\bibfnamefont {B.}~\bibnamefont
  {Huang}}, \bibinfo {author} {\bibfnamefont {G.}~\bibnamefont {Clark}},
  \bibinfo {author} {\bibfnamefont {E.}~\bibnamefont {Navarro-Moratalla}},
  \bibinfo {author} {\bibfnamefont {D.~R.}\ \bibnamefont {Klein}}, \bibinfo
  {author} {\bibfnamefont {R.}~\bibnamefont {Cheng}}, \bibinfo {author}
  {\bibfnamefont {K.~L.}\ \bibnamefont {Seyler}}, \bibinfo {author}
  {\bibfnamefont {D.}~\bibnamefont {Zhong}}, \bibinfo {author} {\bibfnamefont
  {E.}~\bibnamefont {Schmidgall}}, \bibinfo {author} {\bibfnamefont {M.~A.}\
  \bibnamefont {McGuire}}, \bibinfo {author} {\bibfnamefont {D.~H.}\
  \bibnamefont {Cobden}}, \bibinfo {author} {\bibfnamefont {W.}~\bibnamefont
  {Yao}}, \bibinfo {author} {\bibfnamefont {D.}~\bibnamefont {Xiao}}, \bibinfo
  {author} {\bibfnamefont {P.}~\bibnamefont {Jarillo-Herrero}},\ and\ \bibinfo
  {author} {\bibfnamefont {X.}~\bibnamefont {Xu}},\ }\bibfield  {title}
  {\bibinfo {title} {Layer-dependent ferromagnetism in a van der {Waals}
  crystal down to the monolayer limit},\ }\href
  {https://doi.org/10.1038/nature22391} {\bibfield  {journal} {\bibinfo
  {journal} {Nature}\ }\textbf {\bibinfo {volume} {546}},\ \bibinfo {pages}
  {270} (\bibinfo {year} {2017})}\BibitemShut {NoStop}%
\bibitem [{\citenamefont {Gong}\ \emph {et~al.}(2017)\citenamefont {Gong},
  \citenamefont {Li}, \citenamefont {Li}, \citenamefont {Ji}, \citenamefont
  {Stern}, \citenamefont {Xia}, \citenamefont {Cao}, \citenamefont {Bao},
  \citenamefont {Wang}, \citenamefont {Wang}, \citenamefont {Qiu},
  \citenamefont {Cava}, \citenamefont {Louie}, \citenamefont {Xia},\ and\
  \citenamefont {Zhang}}]{gong_discovery_2017}%
  \BibitemOpen
  \bibfield  {author} {\bibinfo {author} {\bibfnamefont {C.}~\bibnamefont
  {Gong}}, \bibinfo {author} {\bibfnamefont {L.}~\bibnamefont {Li}}, \bibinfo
  {author} {\bibfnamefont {Z.}~\bibnamefont {Li}}, \bibinfo {author}
  {\bibfnamefont {H.}~\bibnamefont {Ji}}, \bibinfo {author} {\bibfnamefont
  {A.}~\bibnamefont {Stern}}, \bibinfo {author} {\bibfnamefont
  {Y.}~\bibnamefont {Xia}}, \bibinfo {author} {\bibfnamefont {T.}~\bibnamefont
  {Cao}}, \bibinfo {author} {\bibfnamefont {W.}~\bibnamefont {Bao}}, \bibinfo
  {author} {\bibfnamefont {C.}~\bibnamefont {Wang}}, \bibinfo {author}
  {\bibfnamefont {Y.}~\bibnamefont {Wang}}, \bibinfo {author} {\bibfnamefont
  {Z.~Q.}\ \bibnamefont {Qiu}}, \bibinfo {author} {\bibfnamefont {R.~J.}\
  \bibnamefont {Cava}}, \bibinfo {author} {\bibfnamefont {S.~G.}\ \bibnamefont
  {Louie}}, \bibinfo {author} {\bibfnamefont {J.}~\bibnamefont {Xia}},\ and\
  \bibinfo {author} {\bibfnamefont {X.}~\bibnamefont {Zhang}},\ }\bibfield
  {title} {\bibinfo {title} {Discovery of intrinsic ferromagnetism in
  two-dimensional van der {Waals} crystals},\ }\href
  {https://doi.org/10.1038/nature22060} {\bibfield  {journal} {\bibinfo
  {journal} {Nature}\ }\textbf {\bibinfo {volume} {546}},\ \bibinfo {pages}
  {265} (\bibinfo {year} {2017})}\BibitemShut {NoStop}%
\bibitem [{\citenamefont {Deng}\ \emph {et~al.}(2018)\citenamefont {Deng},
  \citenamefont {Yu}, \citenamefont {Song}, \citenamefont {Zhang},
  \citenamefont {Wang}, \citenamefont {Sun}, \citenamefont {Yi}, \citenamefont
  {Wu}, \citenamefont {Wu}, \citenamefont {Zhu}, \citenamefont {Wang},
  \citenamefont {Chen},\ and\ \citenamefont {Zhang}}]{deng_gate-tunable_2018}%
  \BibitemOpen
  \bibfield  {author} {\bibinfo {author} {\bibfnamefont {Y.}~\bibnamefont
  {Deng}}, \bibinfo {author} {\bibfnamefont {Y.}~\bibnamefont {Yu}}, \bibinfo
  {author} {\bibfnamefont {Y.}~\bibnamefont {Song}}, \bibinfo {author}
  {\bibfnamefont {J.}~\bibnamefont {Zhang}}, \bibinfo {author} {\bibfnamefont
  {N.~Z.}\ \bibnamefont {Wang}}, \bibinfo {author} {\bibfnamefont
  {Z.}~\bibnamefont {Sun}}, \bibinfo {author} {\bibfnamefont {Y.}~\bibnamefont
  {Yi}}, \bibinfo {author} {\bibfnamefont {Y.~Z.}\ \bibnamefont {Wu}}, \bibinfo
  {author} {\bibfnamefont {S.}~\bibnamefont {Wu}}, \bibinfo {author}
  {\bibfnamefont {J.}~\bibnamefont {Zhu}}, \bibinfo {author} {\bibfnamefont
  {J.}~\bibnamefont {Wang}}, \bibinfo {author} {\bibfnamefont {X.~H.}\
  \bibnamefont {Chen}},\ and\ \bibinfo {author} {\bibfnamefont
  {Y.}~\bibnamefont {Zhang}},\ }\bibfield  {title} {\bibinfo {title}
  {Gate-tunable room-temperature ferromagnetism in two-dimensional
  {Fe3GeTe2}},\ }\href {https://doi.org/10.1038/s41586-018-0626-9} {\bibfield
  {journal} {\bibinfo  {journal} {Nature}\ }\textbf {\bibinfo {volume} {563}},\
  \bibinfo {pages} {94} (\bibinfo {year} {2018})}\BibitemShut {NoStop}%
\bibitem [{\citenamefont {Liu}\ \emph {et~al.}(2024)\citenamefont {Liu},
  \citenamefont {Yu},\ and\ \citenamefont {Liu}}]{liu_twisted_2024}%
  \BibitemOpen
  \bibfield  {author} {\bibinfo {author} {\bibfnamefont {Y.}~\bibnamefont
  {Liu}}, \bibinfo {author} {\bibfnamefont {J.}~\bibnamefont {Yu}},\ and\
  \bibinfo {author} {\bibfnamefont {C.-C.}\ \bibnamefont {Liu}},\ }\bibfield
  {title} {\bibinfo {title} {Twisted {Magnetic} {Van} der {Waals} {Bilayers}:
  {An} {Ideal} {Platform} for {Altermagnetism}},\ }\href
  {https://doi.org/10.1103/PhysRevLett.133.206702} {\bibfield  {journal}
  {\bibinfo  {journal} {Physical Review Letters}\ }\textbf {\bibinfo {volume}
  {133}},\ \bibinfo {pages} {206702} (\bibinfo {year} {2024})},\ \bibinfo
  {note} {arXiv:2404.17146 [cond-mat]}\BibitemShut {NoStop}%
\bibitem [{\citenamefont {Ablimit}\ \emph {et~al.}(2018)\citenamefont
  {Ablimit}, \citenamefont {Sun}, \citenamefont {Jiang}, \citenamefont {Wu},
  \citenamefont {Liu},\ and\ \citenamefont {Cao}}]{ablimit_weak_2018}%
  \BibitemOpen
  \bibfield  {author} {\bibinfo {author} {\bibfnamefont {A.}~\bibnamefont
  {Ablimit}}, \bibinfo {author} {\bibfnamefont {Y.-L.}\ \bibnamefont {Sun}},
  \bibinfo {author} {\bibfnamefont {H.}~\bibnamefont {Jiang}}, \bibinfo
  {author} {\bibfnamefont {S.-Q.}\ \bibnamefont {Wu}}, \bibinfo {author}
  {\bibfnamefont {Y.-B.}\ \bibnamefont {Liu}},\ and\ \bibinfo {author}
  {\bibfnamefont {G.-H.}\ \bibnamefont {Cao}},\ }\bibfield  {title} {\bibinfo
  {title} {Weak metal-metal transition in the vanadium oxytelluride
  {Rb}\$\_\{1-{\textbackslash}delta\}\${V}\$\_2\${Te}\$\_2\${O}},\ }\href
  {https://doi.org/10.1103/PhysRevB.97.214517} {\bibfield  {journal} {\bibinfo
  {journal} {Physical Review B}\ }\textbf {\bibinfo {volume} {97}},\ \bibinfo
  {pages} {214517} (\bibinfo {year} {2018})}\BibitemShut {NoStop}%
\bibitem [{\citenamefont {Ghimire}\ \emph {et~al.}(2018)\citenamefont
  {Ghimire}, \citenamefont {Botana}, \citenamefont {Jiang}, \citenamefont
  {Zhang}, \citenamefont {Chen},\ and\ \citenamefont
  {Mitchell}}]{ghimire_large_2018}%
  \BibitemOpen
  \bibfield  {author} {\bibinfo {author} {\bibfnamefont {N.~J.}\ \bibnamefont
  {Ghimire}}, \bibinfo {author} {\bibfnamefont {A.~S.}\ \bibnamefont {Botana}},
  \bibinfo {author} {\bibfnamefont {J.~S.}\ \bibnamefont {Jiang}}, \bibinfo
  {author} {\bibfnamefont {J.}~\bibnamefont {Zhang}}, \bibinfo {author}
  {\bibfnamefont {Y.-S.}\ \bibnamefont {Chen}},\ and\ \bibinfo {author}
  {\bibfnamefont {J.~F.}\ \bibnamefont {Mitchell}},\ }\bibfield  {title}
  {\bibinfo {title} {Large anomalous {Hall} effect in the chiral-lattice
  antiferromagnet {CoNb3S6}},\ }\href
  {https://doi.org/10.1038/s41467-018-05756-7} {\bibfield  {journal} {\bibinfo
  {journal} {Nature Communications}\ }\textbf {\bibinfo {volume} {9}},\
  \bibinfo {pages} {3280} (\bibinfo {year} {2018})}\BibitemShut {NoStop}%
\bibitem [{\citenamefont {Guo}\ \emph {et~al.}(2021)\citenamefont {Guo},
  \citenamefont {Wei}, \citenamefont {Liu}, \citenamefont {Liu},\ and\
  \citenamefont {Lu}}]{guo_eightfold_2021}%
  \BibitemOpen
  \bibfield  {author} {\bibinfo {author} {\bibfnamefont {P.-J.}\ \bibnamefont
  {Guo}}, \bibinfo {author} {\bibfnamefont {Y.-W.}\ \bibnamefont {Wei}},
  \bibinfo {author} {\bibfnamefont {K.}~\bibnamefont {Liu}}, \bibinfo {author}
  {\bibfnamefont {Z.-X.}\ \bibnamefont {Liu}},\ and\ \bibinfo {author}
  {\bibfnamefont {Z.-Y.}\ \bibnamefont {Lu}},\ }\bibfield  {title} {\bibinfo
  {title} {Eightfold {Degenerate} {Fermions} in {Two} {Dimensions}},\ }\href
  {https://doi.org/10.1103/PhysRevLett.127.176401} {\bibfield  {journal}
  {\bibinfo  {journal} {Physical Review Letters}\ }\textbf {\bibinfo {volume}
  {127}},\ \bibinfo {pages} {176401} (\bibinfo {year} {2021})}\BibitemShut
  {NoStop}%
\bibitem [{\citenamefont {Che}\ \emph {et~al.}(2025)\citenamefont {Che},
  \citenamefont {Lv}, \citenamefont {Wu},\ and\ \citenamefont
  {Yang}}]{che_engineering_2025}%
  \BibitemOpen
  \bibfield  {author} {\bibinfo {author} {\bibfnamefont {Y.}~\bibnamefont
  {Che}}, \bibinfo {author} {\bibfnamefont {H.}~\bibnamefont {Lv}}, \bibinfo
  {author} {\bibfnamefont {X.}~\bibnamefont {Wu}},\ and\ \bibinfo {author}
  {\bibfnamefont {J.}~\bibnamefont {Yang}},\ }\bibfield  {title} {\bibinfo
  {title} {Engineering {Altermagnetic} {States} in {Two}-{Dimensional} {Square}
  {Tessellations}},\ }\href {https://doi.org/10.1103/v38b-5by1} {\bibfield
  {journal} {\bibinfo  {journal} {Physical Review Letters}\ }\textbf {\bibinfo
  {volume} {135}},\ \bibinfo {pages} {036701} (\bibinfo {year}
  {2025})}\BibitemShut {NoStop}%
\bibitem [{\citenamefont {Mermin}\ and\ \citenamefont
  {Wagner}(1966)}]{mermin_absence_1966}%
  \BibitemOpen
  \bibfield  {author} {\bibinfo {author} {\bibfnamefont {N.~D.}\ \bibnamefont
  {Mermin}}\ and\ \bibinfo {author} {\bibfnamefont {H.}~\bibnamefont
  {Wagner}},\ }\bibfield  {title} {\bibinfo {title} {Absence of
  {Ferromagnetism} or {Antiferromagnetism} in {One}- or {Two}-{Dimensional}
  {Isotropic} {Heisenberg} {Models}},\ }\href
  {https://doi.org/10.1103/PhysRevLett.17.1133} {\bibfield  {journal} {\bibinfo
   {journal} {Physical Review Letters}\ }\textbf {\bibinfo {volume} {17}},\
  \bibinfo {pages} {1133} (\bibinfo {year} {1966})}\BibitemShut {NoStop}%
\bibitem [{\citenamefont {Gong}\ and\ \citenamefont
  {Zhang}(2019)}]{gong_two-dimensional_2019}%
  \BibitemOpen
  \bibfield  {author} {\bibinfo {author} {\bibfnamefont {C.}~\bibnamefont
  {Gong}}\ and\ \bibinfo {author} {\bibfnamefont {X.}~\bibnamefont {Zhang}},\
  }\bibfield  {title} {\bibinfo {title} {Two-dimensional magnetic crystals and
  emergent heterostructure devices},\ }\href
  {https://doi.org/10.1126/science.aav4450} {\bibfield  {journal} {\bibinfo
  {journal} {Science}\ }\textbf {\bibinfo {volume} {363}},\ \bibinfo {pages}
  {eaav4450} (\bibinfo {year} {2019})}\BibitemShut {NoStop}%
\bibitem [{\citenamefont {Wang}\ \emph {et~al.}(2022)\citenamefont {Wang},
  \citenamefont {Bedoya-Pinto}, \citenamefont {Blei}, \citenamefont {Dismukes},
  \citenamefont {Hamo}, \citenamefont {Jenkins}, \citenamefont {Koperski},
  \citenamefont {Liu}, \citenamefont {Sun}, \citenamefont {Telford},
  \citenamefont {Kim}, \citenamefont {Augustin}, \citenamefont {Vool},
  \citenamefont {Yin}, \citenamefont {Li}, \citenamefont {Falin}, \citenamefont
  {Dean}, \citenamefont {Casanova}, \citenamefont {Evans}, \citenamefont
  {Chshiev}, \citenamefont {Mishchenko}, \citenamefont {Petrovic},
  \citenamefont {He}, \citenamefont {Zhao}, \citenamefont {Tsen}, \citenamefont
  {Gerardot}, \citenamefont {Brotons-Gisbert}, \citenamefont {Guguchia},
  \citenamefont {Roy}, \citenamefont {Tongay}, \citenamefont {Wang},
  \citenamefont {Hasan}, \citenamefont {Wrachtrup}, \citenamefont {Yacoby},
  \citenamefont {Fert}, \citenamefont {Parkin}, \citenamefont {Novoselov},
  \citenamefont {Dai}, \citenamefont {Balicas},\ and\ \citenamefont
  {Santos}}]{wang_magnetic_2022}%
  \BibitemOpen
  \bibfield  {author} {\bibinfo {author} {\bibfnamefont {Q.~H.}\ \bibnamefont
  {Wang}}, \bibinfo {author} {\bibfnamefont {A.}~\bibnamefont {Bedoya-Pinto}},
  \bibinfo {author} {\bibfnamefont {M.}~\bibnamefont {Blei}}, \bibinfo {author}
  {\bibfnamefont {A.~H.}\ \bibnamefont {Dismukes}}, \bibinfo {author}
  {\bibfnamefont {A.}~\bibnamefont {Hamo}}, \bibinfo {author} {\bibfnamefont
  {S.}~\bibnamefont {Jenkins}}, \bibinfo {author} {\bibfnamefont
  {M.}~\bibnamefont {Koperski}}, \bibinfo {author} {\bibfnamefont
  {Y.}~\bibnamefont {Liu}}, \bibinfo {author} {\bibfnamefont {Q.-C.}\
  \bibnamefont {Sun}}, \bibinfo {author} {\bibfnamefont {E.~J.}\ \bibnamefont
  {Telford}}, \bibinfo {author} {\bibfnamefont {H.~H.}\ \bibnamefont {Kim}},
  \bibinfo {author} {\bibfnamefont {M.}~\bibnamefont {Augustin}}, \bibinfo
  {author} {\bibfnamefont {U.}~\bibnamefont {Vool}}, \bibinfo {author}
  {\bibfnamefont {J.-X.}\ \bibnamefont {Yin}}, \bibinfo {author} {\bibfnamefont
  {L.~H.}\ \bibnamefont {Li}}, \bibinfo {author} {\bibfnamefont
  {A.}~\bibnamefont {Falin}}, \bibinfo {author} {\bibfnamefont {C.~R.}\
  \bibnamefont {Dean}}, \bibinfo {author} {\bibfnamefont {F.}~\bibnamefont
  {Casanova}}, \bibinfo {author} {\bibfnamefont {R.~F.~L.}\ \bibnamefont
  {Evans}}, \bibinfo {author} {\bibfnamefont {M.}~\bibnamefont {Chshiev}},
  \bibinfo {author} {\bibfnamefont {A.}~\bibnamefont {Mishchenko}}, \bibinfo
  {author} {\bibfnamefont {C.}~\bibnamefont {Petrovic}}, \bibinfo {author}
  {\bibfnamefont {R.}~\bibnamefont {He}}, \bibinfo {author} {\bibfnamefont
  {L.}~\bibnamefont {Zhao}}, \bibinfo {author} {\bibfnamefont {A.~W.}\
  \bibnamefont {Tsen}}, \bibinfo {author} {\bibfnamefont {B.~D.}\ \bibnamefont
  {Gerardot}}, \bibinfo {author} {\bibfnamefont {M.}~\bibnamefont
  {Brotons-Gisbert}}, \bibinfo {author} {\bibfnamefont {Z.}~\bibnamefont
  {Guguchia}}, \bibinfo {author} {\bibfnamefont {X.}~\bibnamefont {Roy}},
  \bibinfo {author} {\bibfnamefont {S.}~\bibnamefont {Tongay}}, \bibinfo
  {author} {\bibfnamefont {Z.}~\bibnamefont {Wang}}, \bibinfo {author}
  {\bibfnamefont {M.~Z.}\ \bibnamefont {Hasan}}, \bibinfo {author}
  {\bibfnamefont {J.}~\bibnamefont {Wrachtrup}}, \bibinfo {author}
  {\bibfnamefont {A.}~\bibnamefont {Yacoby}}, \bibinfo {author} {\bibfnamefont
  {A.}~\bibnamefont {Fert}}, \bibinfo {author} {\bibfnamefont {S.}~\bibnamefont
  {Parkin}}, \bibinfo {author} {\bibfnamefont {K.~S.}\ \bibnamefont
  {Novoselov}}, \bibinfo {author} {\bibfnamefont {P.}~\bibnamefont {Dai}},
  \bibinfo {author} {\bibfnamefont {L.}~\bibnamefont {Balicas}},\ and\ \bibinfo
  {author} {\bibfnamefont {E.~J.~G.}\ \bibnamefont {Santos}},\ }\bibfield
  {title} {\bibinfo {title} {The {Magnetic} {Genome} of {Two}-{Dimensional} van
  der {Waals} {Materials}},\ }\href {https://doi.org/10.1021/acsnano.1c09150}
  {\bibfield  {journal} {\bibinfo  {journal} {ACS Nano}\ }\textbf {\bibinfo
  {volume} {16}},\ \bibinfo {pages} {6960} (\bibinfo {year}
  {2022})}\BibitemShut {NoStop}%
\bibitem [{\citenamefont {Kopský}\ and\ \citenamefont
  {Litvin}(2002)}]{kopsky_international_2002}%
  \BibitemOpen
  \bibfield  {author} {\bibinfo {author} {\bibfnamefont {V.}~\bibnamefont
  {Kopský}}\ and\ \bibinfo {author} {\bibfnamefont {D.~B.}\ \bibnamefont
  {Litvin}},\ }\href@noop {} {\emph {\bibinfo {title} {International {Tables}
  for {Crystallography} {Volume} {E}: {Subperiodic} groups}}}\ (\bibinfo
  {publisher} {Kluwer Academic Publishers},\ \bibinfo {address} {London},\
  \bibinfo {year} {2002})\BibitemShut {NoStop}%
\bibitem [{\citenamefont {Zeng}\ and\ \citenamefont
  {Zhao}(2024)}]{zeng_description_2024}%
  \BibitemOpen
  \bibfield  {author} {\bibinfo {author} {\bibfnamefont {S.}~\bibnamefont
  {Zeng}}\ and\ \bibinfo {author} {\bibfnamefont {Y.-J.}\ \bibnamefont
  {Zhao}},\ }\bibfield  {title} {\bibinfo {title} {Description of
  two-dimensional altermagnetism: {Categorization} using spin group theory},\
  }\href {https://doi.org/10.1103/PhysRevB.110.054406} {\bibfield  {journal}
  {\bibinfo  {journal} {Physical Review B}\ }\textbf {\bibinfo {volume}
  {110}},\ \bibinfo {pages} {054406} (\bibinfo {year} {2024})}\BibitemShut
  {NoStop}%
\bibitem [{\citenamefont {Zhang}\ \emph {et~al.}(2023)\citenamefont {Zhang},
  \citenamefont {Wu}, \citenamefont {Liu}, \citenamefont {Yu}, \citenamefont
  {Yang},\ and\ \citenamefont {Yao}}]{zhang_encyclopedia_2023}%
  \BibitemOpen
  \bibfield  {author} {\bibinfo {author} {\bibfnamefont {Z.}~\bibnamefont
  {Zhang}}, \bibinfo {author} {\bibfnamefont {W.}~\bibnamefont {Wu}}, \bibinfo
  {author} {\bibfnamefont {G.-B.}\ \bibnamefont {Liu}}, \bibinfo {author}
  {\bibfnamefont {Z.-M.}\ \bibnamefont {Yu}}, \bibinfo {author} {\bibfnamefont
  {S.~A.}\ \bibnamefont {Yang}},\ and\ \bibinfo {author} {\bibfnamefont
  {Y.}~\bibnamefont {Yao}},\ }\bibfield  {title} {\bibinfo {title}
  {Encyclopedia of emergent particles in 528 magnetic layer groups and 394
  magnetic rod groups},\ }\href {https://doi.org/10.1103/PhysRevB.107.075405}
  {\bibfield  {journal} {\bibinfo  {journal} {Physical Review B}\ }\textbf
  {\bibinfo {volume} {107}},\ \bibinfo {pages} {075405} (\bibinfo {year}
  {2023})}\BibitemShut {NoStop}%
\bibitem [{\citenamefont {Cohen}(1993)}]{cohen_course_1993}%
  \BibitemOpen
  \bibfield  {author} {\bibinfo {author} {\bibfnamefont {H.}~\bibnamefont
  {Cohen}},\ }\href {https://doi.org/10.1007/978-3-662-02945-9} {\emph
  {\bibinfo {title} {A {Course} in {Computational} {Algebraic} {Number}
  {Theory}}}},\ \bibinfo {series} {Graduate {Texts} in {Mathematics}}, Vol.\
  \bibinfo {volume} {138}\ (\bibinfo  {publisher} {Springer Berlin
  Heidelberg},\ \bibinfo {address} {Berlin, Heidelberg},\ \bibinfo {year}
  {1993})\BibitemShut {NoStop}%
\bibitem [{\citenamefont {Eick}\ \emph {et~al.}(1997)\citenamefont {Eick},
  \citenamefont {Gähler},\ and\ \citenamefont {Nickel}}]{eick_computing_1997}%
  \BibitemOpen
  \bibfield  {author} {\bibinfo {author} {\bibfnamefont {B.}~\bibnamefont
  {Eick}}, \bibinfo {author} {\bibfnamefont {F.}~\bibnamefont {Gähler}},\ and\
  \bibinfo {author} {\bibfnamefont {W.}~\bibnamefont {Nickel}},\ }\bibfield
  {title} {\bibinfo {title} {Computing {Maximal} {Subgroups} and {Wyckoff}
  {Positions} of {Space} {Groups}},\ }\href
  {https://doi.org/10.1107/S0108767397003462} {\bibfield  {journal} {\bibinfo
  {journal} {Acta Crystallographica Section A Foundations of Crystallography}\
  }\textbf {\bibinfo {volume} {53}},\ \bibinfo {pages} {467} (\bibinfo {year}
  {1997})}\BibitemShut {NoStop}%
\bibitem [{noa()}]{noauthor_notitle_nodate}%
  \BibitemOpen
  \href@noop {} {\bibinfo  {journal} {See Supplemental Material for details}\
  }\BibitemShut {NoStop}%
\bibitem [{\citenamefont {Liu}\ \emph {et~al.}(2023)\citenamefont {Liu},
  \citenamefont {Zhang}, \citenamefont {Yu},\ and\ \citenamefont
  {Yao}}]{liu_msgcorep_2023}%
  \BibitemOpen
\bibfield  {journal} {  }\bibfield  {author} {\bibinfo {author} {\bibfnamefont
  {G.-B.}\ \bibnamefont {Liu}}, \bibinfo {author} {\bibfnamefont
  {Z.}~\bibnamefont {Zhang}}, \bibinfo {author} {\bibfnamefont {Z.-M.}\
  \bibnamefont {Yu}},\ and\ \bibinfo {author} {\bibfnamefont {Y.}~\bibnamefont
  {Yao}},\ }\bibfield  {title} {\bibinfo {title} {{MSGCorep}: {A} package for
  corepresentations of magnetic space groups},\ }\href
  {https://doi.org/10.1016/j.cpc.2023.108722} {\bibfield  {journal} {\bibinfo
  {journal} {Computer Physics Communications}\ }\textbf {\bibinfo {volume}
  {288}},\ \bibinfo {pages} {108722} (\bibinfo {year} {2023})}\BibitemShut
  {NoStop}%
\bibitem [{\citenamefont {James}\ and\ \citenamefont
  {Liebeck}(2001)}]{james_representations_2001}%
  \BibitemOpen
  \bibfield  {author} {\bibinfo {author} {\bibfnamefont {G.~D.}\ \bibnamefont
  {James}}\ and\ \bibinfo {author} {\bibfnamefont {M.~W.}\ \bibnamefont
  {Liebeck}},\ }\href@noop {} {\emph {\bibinfo {title} {Representations and
  characters of groups}}},\ \bibinfo {edition} {2nd}\ ed.\ (\bibinfo
  {publisher} {Cambridge University Press},\ \bibinfo {address} {Cambridge, UK
  ; New York, NY},\ \bibinfo {year} {2001})\BibitemShut {NoStop}%
\bibitem [{\citenamefont {Song}\ \emph {et~al.}(2025)\citenamefont {Song},
  \citenamefont {Yang}, \citenamefont {Jiang}, \citenamefont {Fang},
  \citenamefont {Yang}, \citenamefont {Fang}, \citenamefont {Weng},\ and\
  \citenamefont {Liu}}]{song_constructions_2025}%
  \BibitemOpen
  \bibfield  {author} {\bibinfo {author} {\bibfnamefont {Z.}~\bibnamefont
  {Song}}, \bibinfo {author} {\bibfnamefont {A.~Z.}\ \bibnamefont {Yang}},
  \bibinfo {author} {\bibfnamefont {Y.}~\bibnamefont {Jiang}}, \bibinfo
  {author} {\bibfnamefont {Z.}~\bibnamefont {Fang}}, \bibinfo {author}
  {\bibfnamefont {J.}~\bibnamefont {Yang}}, \bibinfo {author} {\bibfnamefont
  {C.}~\bibnamefont {Fang}}, \bibinfo {author} {\bibfnamefont {H.}~\bibnamefont
  {Weng}},\ and\ \bibinfo {author} {\bibfnamefont {Z.-X.}\ \bibnamefont
  {Liu}},\ }\bibfield  {title} {\bibinfo {title} {Constructions and
  applications of irreducible representations of spin-space groups},\ }\href
  {https://doi.org/10.1103/PhysRevB.111.134407} {\bibfield  {journal} {\bibinfo
   {journal} {Physical Review B}\ }\textbf {\bibinfo {volume} {111}},\ \bibinfo
  {pages} {134407} (\bibinfo {year} {2025})}\BibitemShut {NoStop}%
\bibitem [{\citenamefont
  {Dabbaghian-Abdoly}(2005)}]{dabbaghian-abdoly_algorithm_2005}%
  \BibitemOpen
  \bibfield  {author} {\bibinfo {author} {\bibfnamefont {V.}~\bibnamefont
  {Dabbaghian-Abdoly}},\ }\bibfield  {title} {\bibinfo {title} {An algorithm
  for constructing representations of finite groups},\ }\href
  {https://doi.org/10.1016/j.jsc.2005.01.002} {\bibfield  {journal} {\bibinfo
  {journal} {Journal of Symbolic Computation}\ }\textbf {\bibinfo {volume}
  {39}},\ \bibinfo {pages} {671} (\bibinfo {year} {2005})}\BibitemShut
  {NoStop}%
\bibitem [{noa(2024)}]{noauthor_gap_2024}%
  \BibitemOpen
  \href {https://www.gap-system.org} {\emph {\bibinfo {title} {{GAP} –
  {Groups}, {Algorithms}, and {Programming}, {Version} 4.14.0
  https://www.gap-system.org}}}\ (\bibinfo  {publisher} {The GAP Group},\
  \bibinfo {year} {2024})\BibitemShut {NoStop}%
\end{thebibliography}%

\end{document}